\documentclass[aps,pra,twocolumn,10pt,letterpaper]{revtex4-1}

\usepackage{xcolor, graphicx}
\usepackage{amsmath, amssymb}
\usepackage[colorlinks=true,urlcolor=blue,citecolor=blue,linkcolor=blue]{hyperref}

\begin{document}

\title{Entanglement in macroscopic systems}

\author{J. Sperling}\email{jan.sperling@physics.ox.ac.uk}
\affiliation{Clarendon Laboratory, University of Oxford, Parks Road, Oxford OX1 3PU, United Kingdom}

\author{I. A. Walmsley}
\affiliation{Clarendon Laboratory, University of Oxford, Parks Road, Oxford OX1 3PU, United Kingdom}

\date{\today}

\begin{abstract}
	We present a theoretical study of entanglement in ensembles consisting of an arbitrary number of particles.
	Multipartite entanglement criteria in terms of observables are formulated for a fixed number of particles as well as for systems with a fluctuating particle number.
	To access the quality of the verified entanglement, the operational measure of the entanglement visibility is introduced.
	As an example, we perform an analytical characterization of quantum systems composed of interacting harmonic oscillators and witness the entanglement via energy measurements.
	Our analysis shows that the detectable entanglement decays for macroscopic particle numbers without the need for decoherence processes and for all considered coupling regimes.
	We further study thermal states of the given correlated system together with the temperature dependence of entanglement.
\end{abstract}

\maketitle

\section{Introduction}

	In the early days of quantum physics, entanglement was one of the new concepts that enriched the scientific dispute which addressed the question of whether or not quantum theory could be a valid description of nature \cite{EPR35,S35}.
	Since those days, entanglement has been a subject of many studies in mathematical and applied physics.
	In particular, the question of nonlocality has been considered to falsify classical local-hidden-variable models of quantum physics \cite{B64,CHSH69,BCPSW14}, which has been demonstrated in pioneering experiments \cite{AGR81,ADR82} and recent loophole-free implementations \cite{Hetal15,Getal16,Setal16}.
	Nowadays, quantum properties of systems are no longer limited to purely academic investigations \cite{DM03}.
	Rather they serve as resources for quantum technologies, such as quantum computation and communication \cite{NC00,HHHH09,WPGCRSL12}.
	In addition, entanglement has shown its usefulness in metrology \cite{T12,HLKSWWPS12,M13,OAGKAL16}, and unexpected relations to other fields of physics are currently explored; see, e.g., Refs. \cite{H16,C15}.

	The phenomenon of entanglement requires at least two systems which are quantum correlated---i.e., they are not separable.
	Yet, the richness and complexity of inseparability is only truly recognizable if a manifold of individual particles is jointly considered \cite{AFOV08,HV13,LM13,SSV14,GSVCRTF16}.
	The number of different forms of multipartite entanglement rapidly increases with the size of the system and, therefore, the computational effort to treat entangled many-body quantum systems \cite{ASI04,GSVCRTF15}.
	In particular, a collection of a macroscopic number of particles seems to go beyond the scope of current methods to characterize them.
	Nonetheless, some theoretical and experimental approaches tackle such sophisticated problems \cite{AMJHTLB12,SV13,CMP14,RMJFT14,AWSC15,NFB14,GSVCRTF15,MZHCV15,GSVCRTF16}.

	In addition to the many-particle approach studied in this work, other macroscopic properties of inseparable systems have been investigated \cite{V08}.
	For example, one can consider large distances between or high masses of entangled systems \cite{Uetal07,MRSDC08}, or one infers entanglement at high temperatures \cite{V04,KFCDA15}.
	The latter relation inspired the idea to relate entanglement with thermodynamics \cite{BP08,H08,CS15,GHRRS16}.
	This includes investigations of phase transitions \cite{ON02,OAFF02} and energy differences that arise from quantum correlations \cite{SV13,PHHSBA15}.
	Moreover, a method has been recently introduced which connects a defined notation of macroscopicity to the geometric entanglement of the system \cite{TPKJM16}.
	In addition, coarse graining of measurements has been identified as one source which diminishes entanglement between two macroscopic ensembles \cite{M80,RSS11,WGRS13,SGS14,JLK14}.

	Along with entanglement, other measures of quantumness have been investigated to characterize the classical or quantum origin of correlations; see Ref. \cite{MBCPV12} for a review.
	Quantum correlations beyond entanglement have sometimes a higher importance for applications in quantum information \cite{DSC08} and quantum optics \cite{FP12,ASV13}.
	Also, it has been shown that the structural properties of entanglement are more important resources for quantum computation than its magnitude \cite{DFC05,GFE09}.

	The vast number of studies also demonstrates that entanglement becomes increasingly relevant for different areas of physics.
	However, the detection of this quantum property remains a cumbersome task.
	Among the manifold of proposed entanglement probes \cite{GT09}, the entanglement witness approach is one of the best established ones \cite{HHH96,HHH01}; see also Ref. \cite{BEKGWGHBLS04} for an early experimental application to multipartite systems.
	The construction and optimization of such witnesses has been intensively studied, e.g., in Refs. \cite{LKCH00,T05,HE06,SV09,SRLR17}.
	The witnessing method also allows for the formulation of Bell-like entanglement tests \cite{T00}.
	In Ref. \cite {SV13}, we introduced a systematic technique to formulate multipartite entanglement witnesses.
	The experimental application gave a deeper insight into the complex structure of multimode light fields \cite{GSVCRTF15,GSVCRTF16}.

	In this work, we consider the problem of detecting entanglement in systems with a macroscopic particle number.
	We formulate general methods to witness multipartite entanglement and apply them to a system that consists of interacting harmonic oscillators, including full analytical results.
	In the first step, we analyze entanglement in systems with an arbitrarily large, but fixed number of particles.
	Beyond that, we characterize entanglement in systems which can have a fluctuating number of particles in a second step.
	Especially for the latter case, we derive an approach that allows for the construction of entanglement tests.

	This work is organized as follows.
	In Sec. \ref{sec:preliminaries}, we recapitulate a technique for detecting entanglement which is applied to the case of fixed particle numbers.
	This section also includes the introduction of the notion of a separable spectrum of an operator and the entanglement visibility.
	An ensemble of a fixed number of interacting particles is characterized in Sec. \ref{sec:fixed}.
	Starting from a detailed analysis of the bipartite scenario, we characterize full and partial entanglement as a function of the particle number and the coupling strength.
	In Sec. \ref{sec:Fock}, we construct a method to verify entanglement in ensembles which have a random particle number using the Fock-space formalism.
	It is particularly applied to study the temperature dependence of entanglement in the given, correlated system.
	We conclude in Sec. \ref{sec:Conclusions}.

\section{Preliminaries}\label{sec:preliminaries}

	In this section, we present the basic concepts used.
	Previously formulated entanglement criteria are described in Sec. \ref{subsec:Detection}.
	The operational measure of entanglement is introduced in Sec. \ref{subsec:Visibility}.
	Eventually, the considered physical model is discussed in Sec. \ref{subsec:Model}.

\subsection{Entanglement detection}\label{subsec:Detection}

	We consider an $N$-fold tensor-product Hilbert space $\mathcal H^{\otimes N}$ for a system of $N$ particles.
	We suppose that all individual particles are described by (or embedded into) a continuous-variable Hilbert space $\mathcal H$.
	A pure, fully separable state is a normalized tensor-product vector,
	\begin{align}\label{eq:FullSep}
		|\psi^\mathrm{(sep)}\rangle=|\psi_1\rangle\otimes\dots\otimes|\psi_N\rangle.
	\end{align}
	Mixed, fully separable states $\hat\rho^\mathrm{(sep)}$ are a convex combination of those pure ones \cite{W89},
	\begin{align}\label{eq:FullSepMixed}
		\hat\rho^\mathrm{(sep)}=\int dP(\psi^\mathrm{(sep)})|\psi^\mathrm{(sep)}\rangle\langle\psi^\mathrm{(sep)}|,
	\end{align}
	with $P$ being a probability distribution over the set of pure states \eqref{eq:FullSep}.
	Full separability serves---for most parts of this work---as our fundamental reference for a classically correlated system since any other form of partial entanglement necessarily excludes full separability.

	Independently of the form of multipartite entanglement, witnesses can be used to verify such quantum correlations \cite{HHH01,HHH96}.
	The construction of witnesses can be done in several ways; see, e.g., Refs. \cite{LKCH00,T05,HE06,SV09}.
	The approach we pursue is based on the so-called separability eigenvalue problem \cite{SV13}, which is described in the following.

	A state $\hat\rho$ is entangled if and only if there exists a Hermitian operator $\hat L$ such that
	\begin{align}\label{eq:EntCrit}
		\langle\hat L\rangle_{\hat\rho}<\lambda^\mathrm{(sep)}_{\min},
	\end{align}
	where $\lambda^\mathrm{(sep)}_{\min}$ denotes the minimal expectation value that can be attained for a separable state \cite{T05,SV13}.
	We can assume that $\hat L$ is a positive semidefinite operator \cite{SV09}.
	The left-hand side of inequality \eqref{eq:EntCrit} relates to the measurement that is conducted in an experiment.
	The right-hand side, however, has to be obtained from theory.

	For determining $\lambda^\mathrm{(sep)}_{\min}$, we introduced the separability eigenvalue equations \cite{SV13},
	\begin{align}\label{eq:SepEvalEqs}
		\hat L_{\psi_{1},\ldots,\psi_{j-1},\psi_{j+1},\ldots,\psi_{N}}
		|\psi_j\rangle=\lambda^\mathrm{(sep)}|\psi_j\rangle,
	\end{align}
	for all $1\leq j\leq N$ and partially reduced operators $\hat L_{\psi_{1},\ldots,\psi_{j-1},\psi_{j+1},\ldots,\psi_{N}}$, which are defined as
	\begin{align}\label{eq:PartRedOp}
	\begin{aligned}
		&\hat L_{\psi_{1},\ldots,\psi_{j-1},\psi_{j+1},\ldots,\psi_{N}}
		\\
		=&\mathrm{tr}_1\cdots\mathrm{tr}_{j-1}\mathrm{tr}_{j+1}\cdots\mathrm{tr}_N
		\Big[\hat L
		\\
		&\times\left(
			\bigotimes_{i=1}^{j-1}(|\psi_i\rangle\langle\psi_i|)
			\otimes \hat 1 \otimes
			\bigotimes_{i=j+1}^N(|\psi_i\rangle\langle\psi_i|)
		\right)\Big],
	\end{aligned}
	\end{align}
	where $\hat 1$ is the identity operator of the single-particle space $\mathcal H$.
	The operator \eqref{eq:PartRedOp} is a function of the $N-1$ states $|\psi_i\rangle$ ($i\neq j$), which is indicated by its index, and it maps the $j$th single-particle state $|\psi_j\rangle\in\mathcal H$ to an element of $\mathcal H$, which is used to formulate the separability eigenvalue equations \eqref{eq:SepEvalEqs} \cite{comment2}.
	In addition, the value $\lambda^\mathrm{(sep)}$ in Eq. \eqref{eq:SepEvalEqs} is referred to as the separability eigenvalue, and the normalized product vector $|\psi^\mathrm{(sep)}\rangle=\bigotimes_{i=1}^N|\psi_i\rangle$ is the separability eigenvector.

	We can define the separable spectrum, which is the set of all separability eigenvalues,
	\begin{align}\label{eq:SepSpect}
		\sigma^\mathrm{(sep)}(\hat L)=\{\lambda^\mathrm{(sep)}:\lambda^\mathrm{(sep)}\text{ solves \eqref{eq:SepEvalEqs}}\}.
	\end{align}
	It is similar to the concept of a spectrum $\sigma(\hat L)$ for the standard eigenvalue problem $\hat L|\psi\rangle=\lambda|\psi\rangle$.
	The desired minimal bound $\lambda_{\min}^\mathrm{(sep)}$ in the entanglement condition \eqref{eq:EntCrit} is the minimal separability eigenvalue $\lambda^\mathrm{(sep)}$  \cite{SV13}, i.e., the minimum of the separable spectrum,
	\begin{align}
		\lambda^\mathrm{(sep)}_{\min}=\min\sigma^\mathrm{(sep)}(\hat L).
	\end{align}
	The method of separability eigenvalue equations has been used---beyond its experimental application \cite{GPMMSVP14,GSVCRTF15,GSVCRTF16}---to characterize the emission of semiconductor structures \cite{PFSV12,PFSV13}, to formulate entanglement quasiprobabilities \cite{SV09a,SV12,BSV17}, and to classify quantum channels \cite{BSS16}.

\subsection{Entanglement visibility}\label{subsec:Visibility}

	In order to access the quality of the verified entanglement in terms of inequality \eqref{eq:EntCrit}, we define the entanglement visibility for a state $\hat\rho$ as
	\begin{align}\label{eq:EntVis}
		\mathcal V^\mathrm{(ent)}=
		\frac{\lambda^\mathrm{(sep)}_{\min}-\langle\hat L\rangle_{\hat\rho}}{\lambda^\mathrm{(sep)}_{\min}+\langle \hat L\rangle_{\hat\rho}}.
	\end{align}
	The nominator directly relates to our entanglement criterion \eqref{eq:EntCrit}, and it is invariant under positive scaling and translation transformations of the operator $\hat L$ \cite{comment3}.
	The quantity $\mathcal V^\mathrm{(ent)}$ resembles the visibility or contrast as it is used in optics.
	The larger the value $\mathcal V^\mathrm{(ent)}>0$, the more significant is the entanglement verification in terms of $\hat L$, whose expectation value is then sufficiently well separated from the bound $\lambda^\mathrm{(sep)}_{\min}$; cf. inequality \eqref{eq:EntCrit}.
	A value close to zero indicates that the resolution of a detection device, measuring $\langle\hat L\rangle_{\hat\rho}$, is not sufficient to significantly verify entanglement.
	In addition, $\mathcal V^\mathrm{(ent)}\leq 0$ means that no entanglement could be detected via the observable $\hat L$.

	The quantity $\mathcal V^\mathrm{(ent)}$ serves as an operational measure of entanglement \cite{SV11a}; i.e., an unsuccessful test ($\mathcal V^\mathrm{(ent)}\leq 0$) does not imply separability since another choice of $\hat L$ might identify entanglement.
	Its operational meaning relates to a specific, performed measurement of the observable $\hat L$.
	Yet, if one maximizes this contrast over the operators $\hat L$, then $\mathcal V^\mathrm{(ent)}$ can be related to witness-based entanglement monotones \cite{B05,BV06}.
	The here-defined entanglement visibility has been recently applied to characterize the emission of entangled light from a microcavity in an optical resonator pumped by a frequency comb \cite{PF17}.

	The observable $\hat L$ can detect entanglement via condition \eqref{eq:EntCrit} if and only if the eigenspace to the minimal eigenvalue contains no separable state \cite{SV11}.
	Suppose $\lambda_{\min}=\min\sigma(\hat L)$ is the minimal standard eigenvalue of $\hat L$.
	Then $\lambda_{\min}<\lambda_{\min}^\mathrm{(sep)}$ is required to ensure that inequality \eqref{eq:EntCrit} can be satisfied in principle.
	We can also conclude that the maximal visibility \eqref{eq:EntVis} is attained for states which are eigenvectors to the minimal eigenvalue,
	\begin{align}\label{eq:MaxVis}
		\mathcal V^\mathrm{(ent)}_{\max}=\max_{\hat\rho}\{\mathcal V^\mathrm{(ent)}\}=\frac{\lambda^\mathrm{(sep)}_{\min}-\lambda_{\min}}{\lambda^\mathrm{(sep)}_{\min}+\lambda_{\min}}\geq\mathcal V^\mathrm{(ent)},
	\end{align}
	because the expectation value of $\hat L$ is $\lambda_{\min}$ for those states and $\mathcal V^\mathrm{(ent)}$ decreases for increasing expectation values.

\subsection{System of interacting oscillators}\label{subsec:Model}

\begin{figure}[t]
	\includegraphics[width=0.3\textwidth]{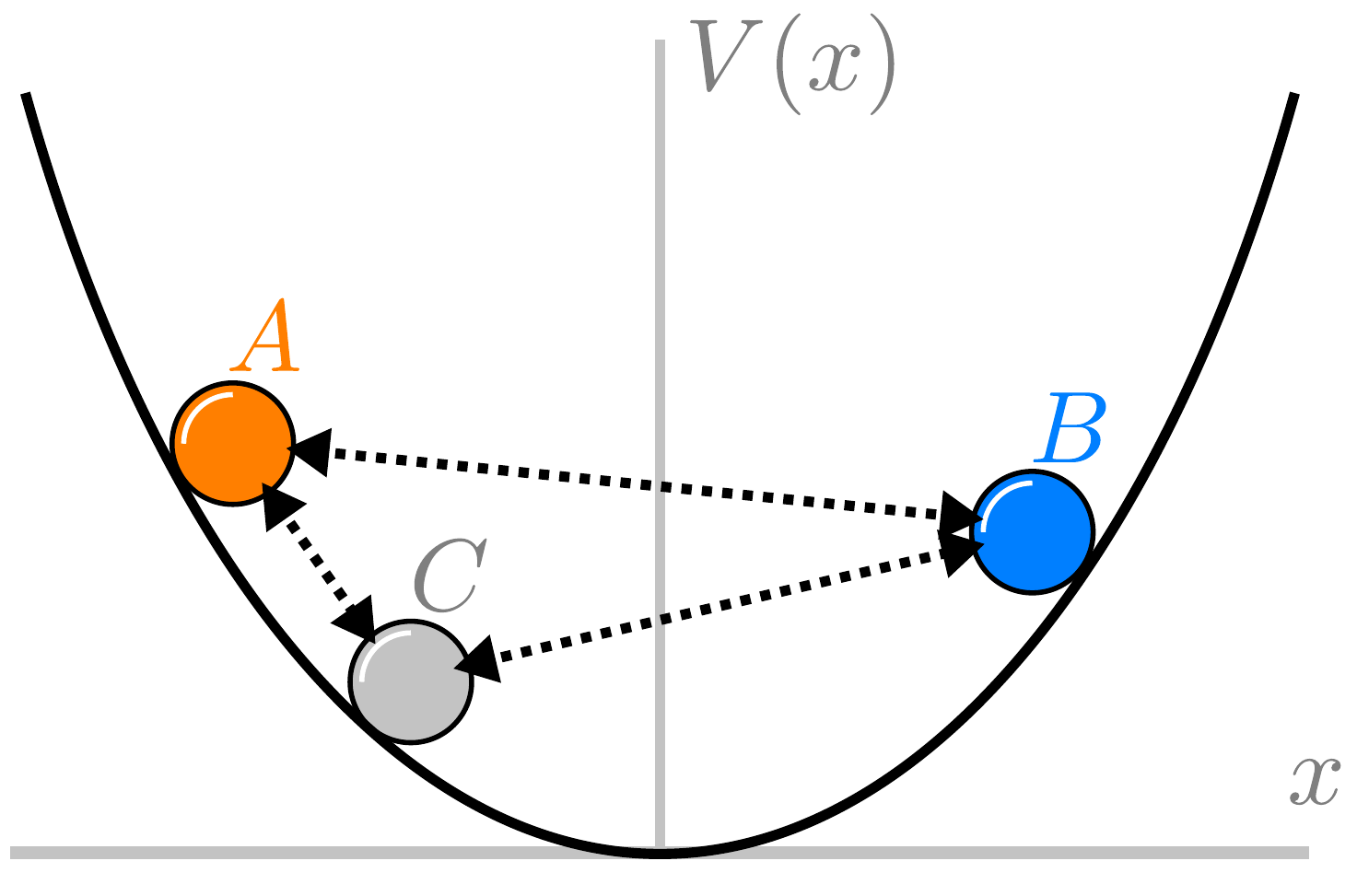}
	\caption{(Color online)
		Schematic sketch of the model for $N=3$ particles ($A$, $B$, and $C$).
		They are trapped in an external potential $V(x)\propto x^2$.
		Each of the particles interacts with any other (dashed arrows), which is described with a force $F_{j\to i}=-\kappa(x_i-x_j)$, which acts on particle $i$ due to the influence of particle $j$ ($i,j\in\{A,B,C\}$).
	}\label{fig:drawing1}
\end{figure}

	Now, we introduce our physical model that is particularly characterized in this work (see Fig. \ref{fig:drawing1}).
	We consider a system of $N$ particles with identical masses $m$ which propagate in an external, one-dimensional potential $V(x)$.
	A second-order expansion of the potential around a minimum ($x=0$) yields a quadratic function, $V(x)\propto x^2$.
	In addition, all particles interact pairwise with each other.
	The strength of this interaction depends on the distance between the particles, $|x_i-x_j|$.
	In first-order approximation, this gives a linear force and a quadratic interaction contribution to the energy.

	In conclusion, the $N$-particle Hamiltonian of the considered system reads
	\begin{align}\label{eq:Hamiltonian}
		\hat H=\sum_{i=1}^N \left(-\frac{\hbar^2\partial_{x_i}^2}{2m}+\frac{m\Omega^2 x_i^2}{2}\right)+\frac{\kappa}{4}\sum_{i,j=1}^N(x_i-x_j)^2.
	\end{align}
	Here $\kappa\geq0$ is the coupling constant and $\Omega$ is the eigenfrequency of the external potential.
	The first sum in this Hamiltonian presents the local Hamiltonian for each individual particle.
	The second sum describes the interaction part between the particles, which induces the entanglement as we demonstrate; see also Ref. \cite{AW08}.

	To quantify the relative interaction strength, we can introduce the ratio $\mathcal R$ between the interaction parameter $\kappa$ and the coupling $m\Omega^2$ to the external potential,
	\begin{align}\label{eq:CouplingRatio}
		\mathcal R=\frac{\kappa}{m\Omega^2}\geq0.
	\end{align}
	A value $\mathcal R=0$ corresponds to a noninteracting system.
	The region $\mathcal R\ll 1$ (i.e., a value close to zero) defines the weak-coupling regime.
	The balanced and strong-coupling cases are represented by $\mathcal R\approx 1$ and $\mathcal R\gg 1$, respectively.

	In addition, in our system of coupled harmonic oscillators \eqref{eq:Hamiltonian}, some characteristic dimensions emerge for the position $x$, the energy $\mathcal E$, and---for Sec. \ref{sec:Fock}---the temperature $T$.
	Those basic units of the system are
	\begin{align}\label{eq:NaturalUnits}
		u_x=\sqrt{\frac{\hbar}{m\Omega}},
		\text{ }
		u_\mathcal{E}=\hbar\Omega,
		\text{ and }
		u_T=\frac{u_\mathcal{E}}{k},
	\end{align}
	where $k$ denotes the Boltzmann constant.
	These units will serve as our scaling parameters throughout this work.

	It is also worth mentioning that the given system consists of $N$ distinguishable particles.
	Entanglement in systems of indistinguishable particles can be also treated \cite{GKM11}.
	We derived the corresponding entanglement conditions together with a modified version of the separability eigenvalue equations, which takes the exchange symmetry of bosons and fermions into account, and compared the resulting forms of entanglement \cite{RSV15}.
	Hence, for our fundamental studies in this work, it is sufficient to focus on the case of distinguishable particles.

\section{Entanglement for fixed numbers of particles}\label{sec:fixed}

	In this section, we use the introduced methods to characterize the system under study.
	We start with a two-particle entanglement in Sec. \ref{subsec:Bipart}.
	In Sec. \ref{subsec:Multipart}, we then analyze the entanglement in the many-particle case.
	In addition, we consider partial entanglement in Sec. \ref{subsec:PartEnt}.

\subsection{Two-particle entanglement}\label{subsec:Bipart}

	In this first step, we characterize the entanglement of the bipartite system, $N=2$, which gives some first hints towards the multipartite entanglement properties.
	Here, the subsystems are labeled as $A$ and $B$.
	We apply the entanglement condition \eqref{eq:EntCrit} using the Hamiltonian \eqref{eq:Hamiltonian}, $\hat L=\hat H$, by solving the separability eigenvalue equations \eqref{eq:SepEvalEqs} for this observable (Appendix \ref{app:FullSepEvalEqsHamiltonian}).

	Let us recall some properties of a noninteracting system, which can be described by a vanishing coupling ratio \eqref{eq:CouplingRatio}, $\mathcal R=0$.
	The time-independent Schr\"odinger equation---the eigenvalue problem $\hat H|\psi\rangle=\mathcal E|\psi\rangle$---of the Hamiltonian \eqref{eq:Hamiltonian} can be solved in that case by separation of variables.
	The eigenstates are products of Hermite functions $h^{(n)}$,
	\begin{align}\label{eq:BipartUncorrStates}
		\psi^{(\mathcal R=0)}(x_A,x_B)=h^{(n_A)}\left(\frac{x_A}{u_x}\right)h^{(n_B)}\left(\frac{x_B}{u_x}\right),
	\end{align}
	and the eigenvalues are given by the sum
	\begin{align}\label{eq:BipartUncorrValues}
		\mathcal E^{(\mathcal R=0)}=\hbar\Omega\left(n_A+\frac{1}{2}\right)+\hbar\Omega\left(n_B+\frac{1}{2}\right),
	\end{align}
	for $n_A,n_B\in\mathbb N$ and using the unit $u_x$ in Eq. \eqref{eq:NaturalUnits}; cf. also Appendix \ref{app:HermiteFct}.
	Without interaction, the total system is an uncorrelated ensemble of two particles.

	For interacting particles, $\mathcal R>0$, we get the general, exact solutions in Appendix \ref{app:FullSepEvalEqsHamiltonian}.
	The eigenfunctions and eigenvalues take the forms
	\begin{align}\label{eq:BipartEstates}
	\begin{aligned}
		&\psi(x_A,x_B)
		\\=&h^{(n_\parallel)}\left(\frac{x_A+x_B}{\sqrt 2 u_x}\right)
		h^{(n_\perp)}\left(\sqrt[4]{1+2\mathcal R}\frac{x_A-x_B}{\sqrt 2 u_x}\right),
	\end{aligned}
	\end{align}
	where $n_\parallel,n_\perp\in\mathbb N$, and
	\begin{align}\label{eq:BipartEsvalues}
		\mathcal E=\hbar\Omega \left(n_\parallel+\frac{1}{2}\right)
		+\hbar\Omega\sqrt{1+2\mathcal R}\left(n_\perp+\frac{1}{2}\right),
	\end{align}
	respectively.
	The index ``$\parallel$'' or ``$\perp$'' label the contribution which is parallel or perpendicular to $(1,1)^\mathrm{T}$ in the $x_A$-$x_B$-plane, respectively.
	In contrast to these solutions of the standard eigenvalue problem, the separability eigenfunctions and separability eigenvalues are
	\begin{align}\label{eq:BipartEsepstates}
	\begin{aligned}
		&\psi^\mathrm{(sep)}(x_A,x_B)=\psi_A(x_A)\psi_B(x_B)
		\\=&h^{(n_A)}\left(\sqrt[4]{1+\mathcal R}\frac{x_A}{u_x}\right)
		h^{(n_B)}\left(\sqrt[4]{1+\mathcal R}\frac{x_B}{u_x}\right)
	\end{aligned}
	\end{align}
	and
	\begin{align}\label{eq:BipartEsepvalues}
		\mathcal E^\mathrm{(sep)}=\hbar\Omega\sqrt{1+\mathcal R}\left(n_A{+}\frac{1}{2}\right)
		+\hbar\Omega\sqrt{1+\mathcal R}\left(n_B{+}\frac{1}{2}\right),
	\end{align}
	respectively (Appendix \ref{app:FullSepEvalEqsHamiltonian}).
	Note that the separability eigenstates \eqref{eq:BipartEsepstates} and eigenvalues \eqref{eq:BipartEsepvalues} are not simply the solutions in Eqs. \eqref{eq:BipartUncorrStates} and \eqref{eq:BipartUncorrValues} for the uncorrelated case.

\begin{figure*}
	\includegraphics[width=\textwidth]{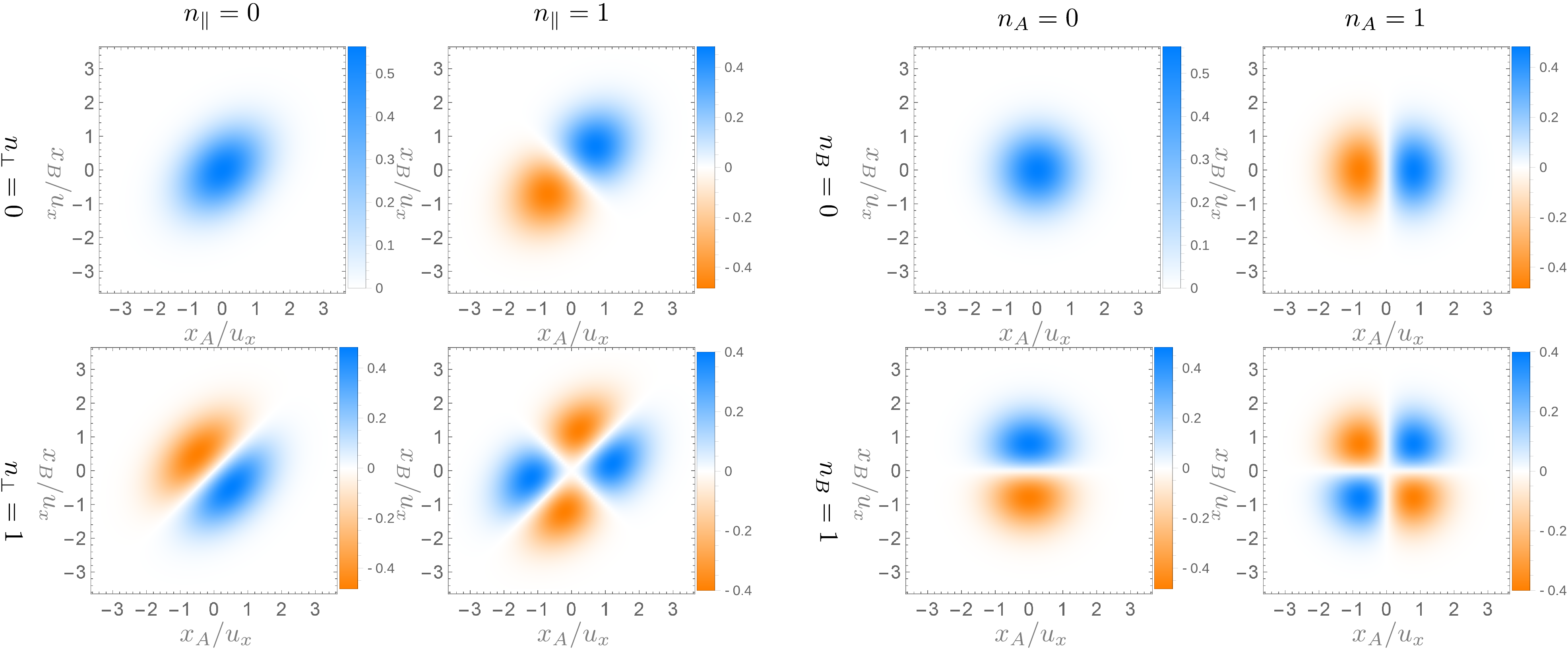}
	\caption{(Color online)
		Left:
		First four eigenstates \eqref{eq:BipartEstates} of the time-independent Schr\"odinger equation.
		The entanglement of those states is directly highlighted by the symmetries along the diagonal and antidiagonal axes.
		Right:
		First four separability eigenstates \eqref{eq:BipartEsepstates} of the solutions of the separability eigenvalue problem of the Hamiltonian $\hat H$.
		Those product states exhibit a symmetric behavior with respect to the horizontal and vertical axes.
		In both cases, we use a coupling ratio $\mathcal R=1.5$.
	}\label{fig:BipartR1dot5}
\end{figure*}

	In Fig. \ref{fig:BipartR1dot5}, eigenfunctions \eqref{eq:BipartEstates} and separability eigenfunctions \eqref{eq:BipartEsepstates} are compared in the balanced interaction regime, $\mathcal R\approx 1$.
	The correlations between the subsystems $A$ and $B$ are visible for the standard eigenstates (Fig. \ref{fig:BipartR1dot5}, panel), because the symmetry axes are parallel and perpendicular to $(1,1)^\mathrm{T}$---the diagonal direction in the $x_A$-$x_B$ plane.
	In contrast, the separability eigenfunctions (Fig. \ref{fig:BipartR1dot5}, right) have the symmetry axes $x_A$ and $x_B$, which is a result of their product structure [Eq. \eqref{eq:BipartEsepstates}].
	Moreover, the entangled states \eqref{eq:BipartEstates} are antisqueezed and squeezed in the directions parallel and perpendicular to $(1,1)^\mathrm{T}$, respectively, in comparison with their separable counterparts \eqref{eq:BipartEsepstates};
	see the scaling of their arguments, $1\leq\sqrt[4]{1+\mathcal R}\leq\sqrt[4]{1+2\mathcal R}$.
	The Gaussian wave functions with the lowest energies, $n_\parallel=0=n_\perp$ and $n_A=0=n_B$, are the Einstein-Podolsky-Rosen-entangled ground state and the uncorrelated ``separable ground state''.

\begin{figure}[t]
	\includegraphics[width=0.25\textwidth]{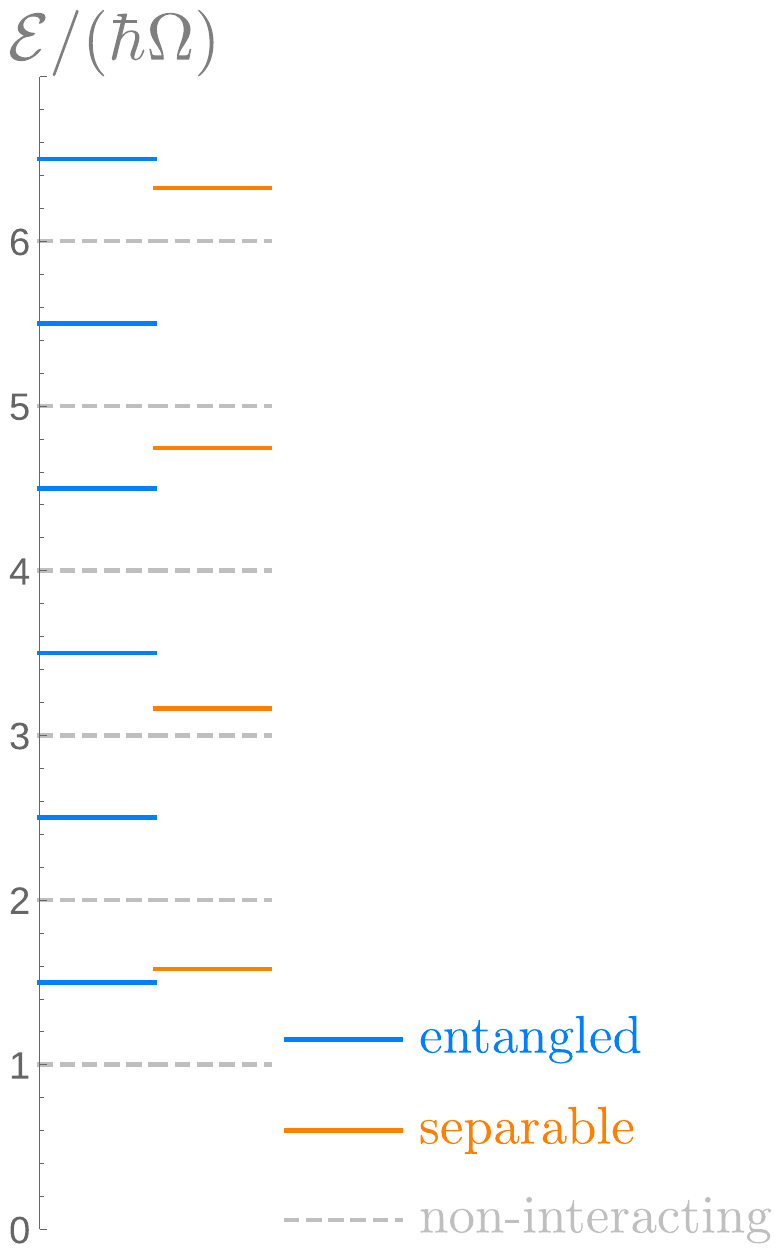}
	\caption{(Color online)
		Energy scheme for a balanced interaction scenario, $\mathcal R=1.5$ (entanglement left and separability right).
		For comparison, the energy spectrum without interaction, $\mathcal R=0$, is also depicted (dashed lines).
	}\label{fig:BipartEnergy}
\end{figure}

	In addition, we plotted the first elements of the energy spectrum $\sigma(\hat H)$ and the separable energy spectrum $\sigma^\mathrm{(sep)}(\hat H)$ [cf. Eq. \eqref{eq:SepSpect}] in Fig. \ref{fig:BipartEnergy}.
	The standard energy levels (left) are smaller in terms of absolute values and have a smaller difference between them when comparing with the case of separability (right).
	In both cases, entanglement and separability, the energy from the interaction, $\mathcal R>0$, leads to an increase of the ground-state energy compared to the case $\mathcal R=0$ (dashed lines).
	The difference of the energies of the ground state and the separability ground state is negative, $\Delta\mathcal E=\mathcal E_{\min}-\mathcal E_{\min}^\mathrm{(sep)}<0$.
	This means that entanglement relates to a diminished ground-state energy of $|\Delta\mathcal E|$, which is the basis of the entanglement criterion \eqref{eq:EntCrit} applied to the Hamiltonian.

\begin{figure}[t]
	\includegraphics[width=0.38\textwidth]{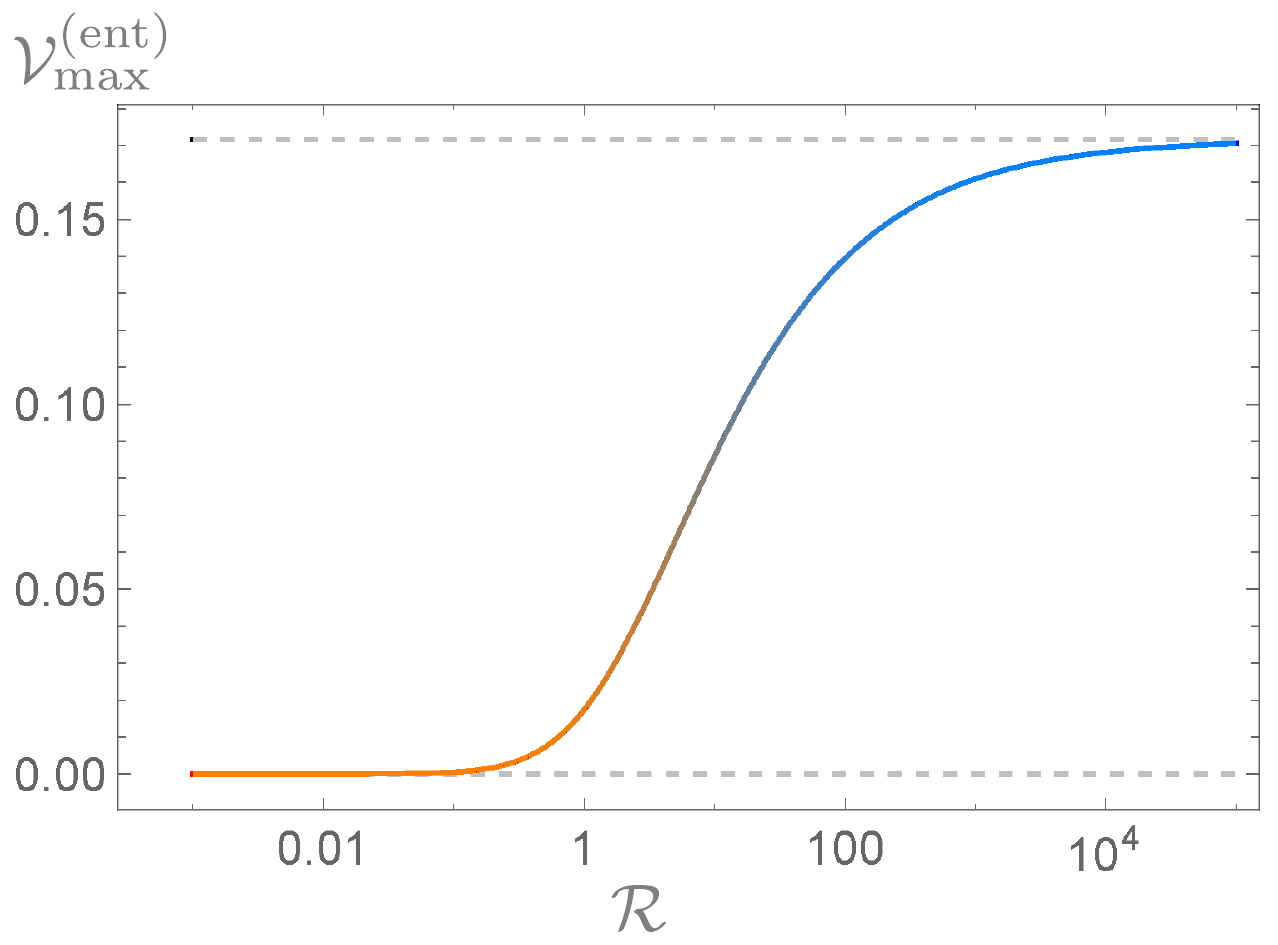}
	\caption{(Color online)
		The maximal entanglement visibility $\mathcal V^\mathrm{(ent)}_{\max}$ as a function of the interaction strength $\mathcal R$.
		Entanglement is verified, $\mathcal V^\mathrm{(ent)}_{\max}>0$, for all $\mathcal R>0$.
		The dashed lines depict the limits for $\mathcal R\to 0$ and $\mathcal R\to\infty$.
	}\label{fig:BipartRdep}
\end{figure}

	Let us now quantify the entanglement of the system in terms of the entanglement visibility \eqref{eq:EntVis}, which is maximized for the entangled ground state,
	\begin{align}
		\mathcal V^\mathrm{(ent)}_{\max}=\frac{\sqrt{1+\mathcal R}-\frac{1}{2}(1+\sqrt{1+2\mathcal R})}{\sqrt{1+\mathcal R}+\frac{1}{2}(1+\sqrt{1+2\mathcal R})},
	\end{align}
	and shown in Fig. \ref{fig:BipartRdep}.
	From $\partial_{\mathcal R}\mathcal V^\mathrm{(ent)}_{\max}>0$ it follows that this visibility is a monotonically increasing function with the minimum $\mathcal V^\mathrm{(ent)}_{\max}=0$ for $\mathcal R=0$.
	For a weak coupling, the entanglement visibility is quite small, which means that the employed device to measure the energy has to be able to resolve small energy differences $\Delta\mathcal E$ to significantly verify entanglement.
	Let us stress that the operational meaning of $\mathcal V^\mathrm{(ent)}$ is based here on the measurement of the total energy, which is given by the two-particle Hamiltonian in Eq. \eqref{eq:Hamiltonian} for $N=2$.
	Finally, for diverging interaction rations, we approach $\lim_{\mathcal R\to\infty}\mathcal V^\mathrm{(ent)}_{\max}=3-2\sqrt 2$ in Fig. \ref{fig:BipartRdep}.

\subsection{Macroscopic entanglement}\label{subsec:Multipart}

	Now, we consider arbitrary particle numbers $N$.
	From Appendix \ref{app:FullSepEvalEqsHamiltonian}, we get the separability eigenvalues
	\begin{align}\label{eq:SepMacroE}
		\mathcal E^\mathrm{(sep)}=\hbar\Omega \sqrt{1+(N-1)\mathcal R}\sum_{j=1}^{N}\left(n_j+\frac{1}{2}\right),
	\end{align}
	where $n_j$ is the excitation number of the $j$th mode, as well as the (standard) eigenvalues
	\begin{align}\label{eq:MacroE}
		\mathcal E
		=\hbar\Omega\left(n_\parallel{+}\frac{1}{2}\right)
		{+}\hbar\Omega\sqrt{1{+}N\mathcal R}\sum_{i=1}^{N-1}\left(n_{i,\perp}{+}\frac{1}{2}\right),
	\end{align}
	where the integers $n_{\parallel}$ denote the excitations along the axis $(1,\dots,1)^\mathrm{T}$ and $n_{i,\perp}$ are the excitations of the $N-1$ perpendicular directions to the axis $(1,\dots,1)^\mathrm{T}$ of the multimode position space $(x_1,\ldots,x_N)^\mathrm{T}$.

\begin{figure}[b]
	\includegraphics[width=0.42\textwidth]{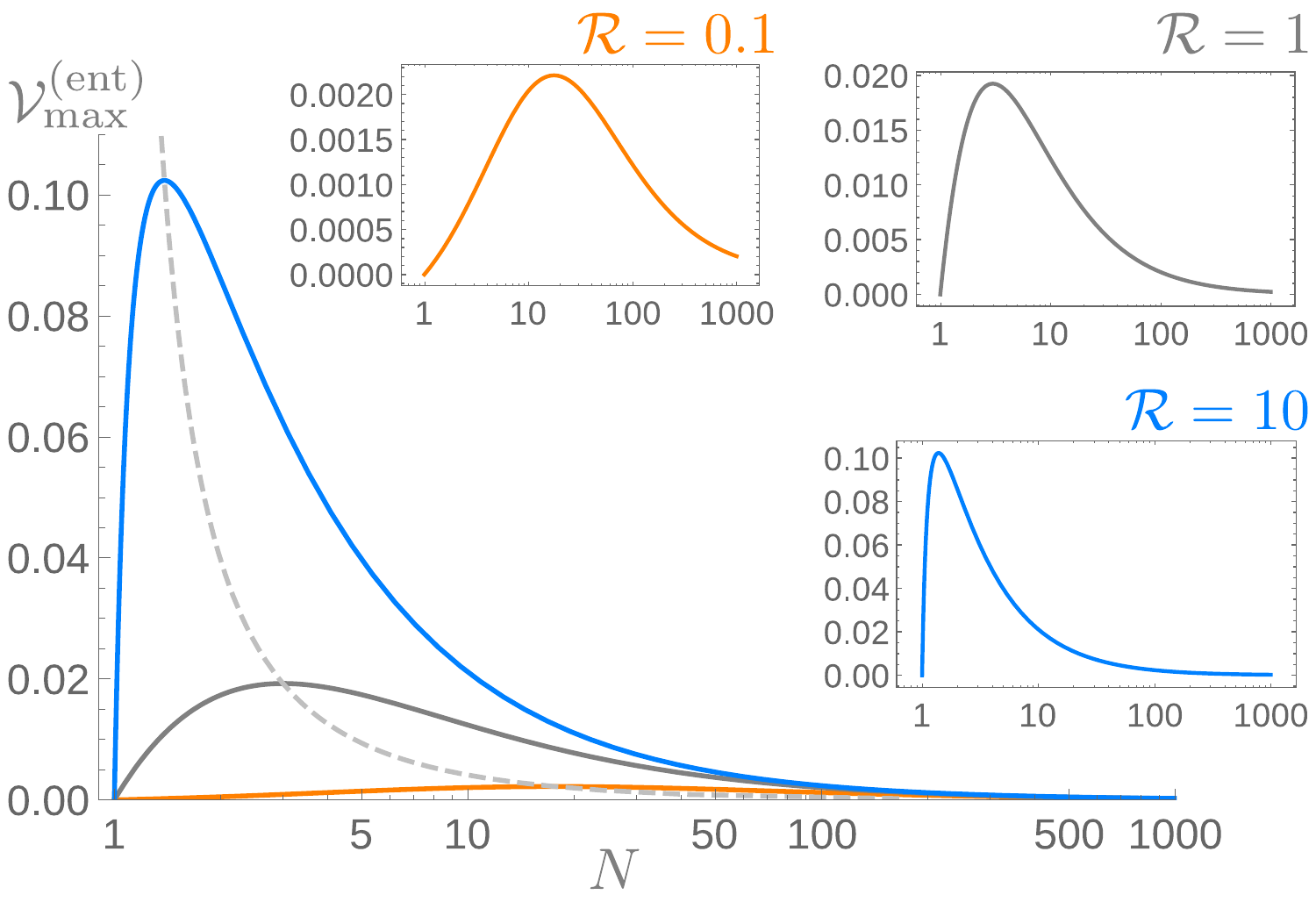}
	\caption{(Color online)
		The visibility $\mathcal V^\mathrm{(ent)}_{\max}$ [Eq. \eqref{eq:maxvisFixed}] is shown as a function of the number of particles, $1\leq N\leq 1\,000$.
		Strong ($\mathcal R=10$, top, blue line), balanced ($\mathcal R=1$, middle, dark gray line), and weak ($\mathcal R=0.1$, bottom, orange line) interaction scenarios are depicted.
		Additionally, the curves are individually plotted in the insets.
		$\mathcal V^\mathrm{(ent)}_{\max}$ for the optimal particle number \eqref{eq:OptPartNum} is shown as a dashed, light gray line.
	}\label{fig:drawing2}
\end{figure}

	The resulting maximal entanglement visibility of the ground state ($n_{\parallel}=0$ and $n_{i,\perp}=0$ for all $i$) of the Hamiltonian \eqref{eq:Hamiltonian} is given by
	\begin{align}\label{eq:maxvisFixed}
		\mathcal V_{\max}^\mathrm{(sep)}
		=\frac{N\sqrt{1{+}(N{-}1)\mathcal R}{-}\big(1{+}(N{-}1)\sqrt{1{+}N\mathcal R}\big)}{N\sqrt{1{+}(N{-}1)\mathcal R}{+}\big(1{+}(N{-}1)\sqrt{1{+}N\mathcal R}\big)}.
	\end{align}
	It is worth pointing out that the minimal energy \eqref{eq:MacroE} and the minimal energy for separable states \eqref{eq:SepMacroE} have the same asymptotic and diverging behavior, $(\hbar\Omega/2)N\sqrt{1+N\mathcal R}$ for $N\gg 1$.
	In Fig. \ref{fig:drawing2}, we show this visibility for different interaction scenarios as a function of the number of particles.
	We study the strong interaction, $\mathcal R\gg1$, weak interaction, $\mathcal R\ll1$, and balanced interaction regime, $\mathcal R\approx 1$.
	Note that noninteger $N$ values are interpolated by directly inserting those numbers into formula \eqref{eq:maxvisFixed}.
	This point is discussed in more detail in Sec. \ref{sec:Fock}.

	One can see in Fig. \ref{fig:drawing2} that a stronger coupling yields a higher entanglement visibility.
	That is, curves for larger $\mathcal R$ values are above those for smaller ones.
	Yet, the distributions for smaller $\mathcal R$ have a larger width than those for stronger coupling (insets in Fig. \ref{fig:drawing2}).
	This means that the visible entanglement for weak coupling is less vulnerable to a change of the particle number.
	In the noninteracting limit, we get $\lim_{\mathcal R\to0}\mathcal V^\mathrm{(ent)}_{\max}=0$ as the ground state becomes factorizable.
	Also, even for an infinite interaction strength the entanglement visibility is bounded, $\lim_{\mathcal R\to\infty}\mathcal V^\mathrm{(ent)}_{\max}=(\sqrt{N}-\sqrt{N-1})^2$.

	In addition, one can optimize the entanglement visibility over the number of particles for a given coupling ratio $\mathcal R$.
	From $0=\partial_N\mathcal V_{\max}^\mathrm{(ent)}$ and after some standard algebra, we get three solutions.
	The one which is physical is
	\begin{align}\label{eq:OptPartNum}
		N_\mathrm{opt}=\frac{1+2\mathcal R+\sqrt{5+4\mathcal R}}{2\mathcal R}.
	\end{align}
	Hence, we can predict the number of particles for a given interaction strength which yields an optimal entanglement visibility (dashed curve in Fig. \ref{fig:drawing2}).

	Further on, we observe---independently of the coupling regime---a decay of the entanglement visibility to zero in the limit of a macroscopic particle number,
	\begin{align}\label{eq:EntDecay}
		\lim_{N\to\infty}\mathcal V^\mathrm{(ent)}_{\max}=0;
	\end{align}
	cf. also Fig. \ref{fig:drawing2}.
	This means that no entanglement can be detected with our energy measurement in this limit.
	Let us emphasize the following two facts.
	First, for all finite $N$ values, the visibility is greater than zero, which proves that entanglement is present.
	Second, the decay of observable entanglement in the macroscopic limit [Eq. \eqref{eq:EntDecay}] happens without the need for employing an additional decoherence mechanism \cite{Z03,S05}.

\subsection{Partial entanglement}\label{subsec:PartEnt}

	In addition, let us also consider partial entanglement for a system consisting of $N$ particles.
	One can collect those particles in $K$ subsystems, which consist of $N_j$ particles for $j=1,\ldots,K$.
	A pure, partially separable state takes the form
	\begin{align}
		|\psi^{(N_1,\dots,N_K)}\rangle=|\psi_1^{(N_1)}\rangle\otimes\cdots\otimes|\psi_{K}^{(N_k)}\rangle,
	\end{align}
	where $|\psi_j^{(N_j)}\rangle\in\mathcal H^{\otimes N_j}$ is an arbitrary state in the $j$th, $N_j$-particle subsystem.
	Mixed, partially separable states are elements of the convex hull of pure state density operators, $|\psi^{(N_1,\dots,N_K)}\rangle\langle\psi^{(N_1,\dots,N_K)}|$---similarly to the case of full separability in Eq. \eqref{eq:FullSepMixed}.

\begin{figure}[b]
	\includegraphics[width=0.38\textwidth]{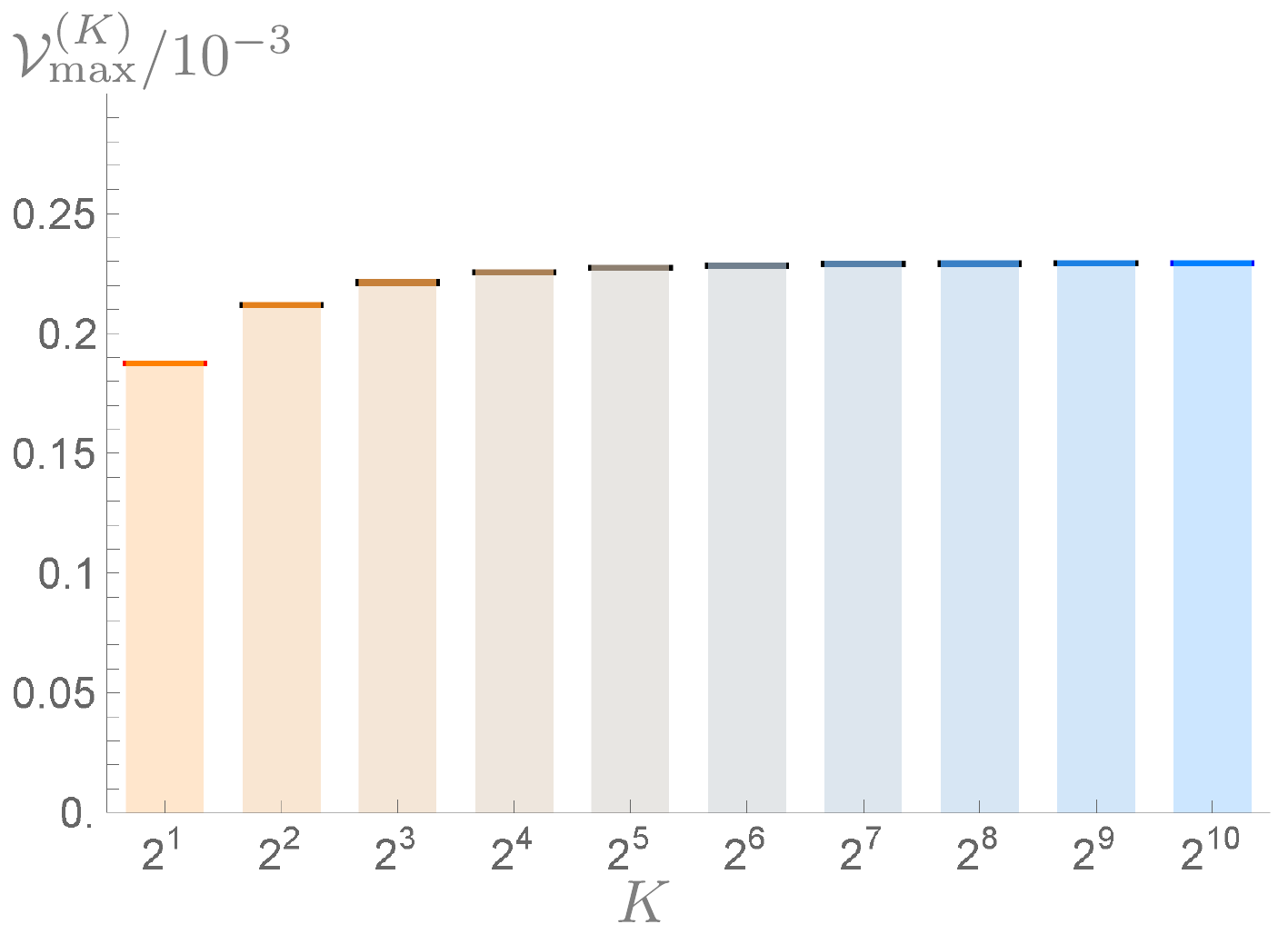}
	\caption{(Color online)
		Maximal entanglement visibility \eqref{eq:PartSepSpectr} for a balanced coupling ($\mathcal R=1$), $N=2^{10}$ particles, and a $K$ partition with $N_1=\dots=N_K=N/K$.
	}\label{fig:Kpart}
\end{figure}

	The minimal energy for such a $K$ partition is given by
	\begin{align}\label{eq:PartSepSpectr}
	\begin{aligned}
		\mathcal E_{\min}^{(N_1,\dots,N_K)}
		=&\frac{\hbar\Omega}{2}\sum_{j=1}^K\sqrt{1+(N-N_j)\mathcal R}
		\\&+\frac{\hbar\Omega}{2}\sum_{j=1}^K(N_j-1)\sqrt{1+N\mathcal R};
	\end{aligned}
	\end{align}
	see Appendix \ref{app:FullSepEvalEqsHamiltonian}.
	Figure \ref{fig:Kpart} shows the maximal entanglement visibility for partial entanglement,
	\begin{align}
		\mathcal V^{(K)}_{\max}=\frac{\mathcal E^{(N_1,\dots,N_K)}_{\min}-\mathcal E_{\min}}{\mathcal E^{(N_1,\dots,N_K)}_{\min}+\mathcal E_{\min}},
	\end{align}
	for $N=1024$ particles.
	The bipartition ($K=2$), consisting of $512$ particles in each subensemble, has the smallest visibility.
	The highest value is attained for the full partition, $K=1024$.
	This results from the fact that a splitting of a given $K$ partition into a finer one, which increases $K$, implies that an inseparable state in the original partition is also entangled in the finer one \cite{SSV14}.
	This proof of principle shows that we are able to infer partial entanglement beyond bipartitions and full inseparability.

\section{Entanglement for unknown particle numbers}\label{sec:Fock}

	From the practical point of view, one cannot truly distinguish, for example, the case of $N$ from the case of $N\pm1$ particles for $N\gg1$.
	Thus, in order to consider a system with such an unknown number of particles, we have to go beyond the restriction of fixed particle numbers.
	In general, a consistent quantum description of systems with an arbitrary number of particles is given in terms of Fock spaces---representing the second quantization.
	An introduction to entanglement in this second quantization can be found in Ref. \cite{V03}, and a comprehensive comparison between the first and second quantization (for light fields) is conducted in Ref. \cite{SR07}.

	Let us emphasize some differences of our approach to typically assumed scenarios.
	The treatment of a fixed number of partitions is the traditional ansatz when studying entanglement.
	Also, in the second quantization, one typically considers entanglement between two or more ensembles of different kinds of particles.
	Or, as it is done in quantum optics, the entanglement is determined between multiple distinct optical modes, which consists of an arbitrary number of photons per mode.
	By contrast, we study the entanglement of particles within one ensemble to characterize the quantum correlations.

	In Sec. \ref{subsec:Fock2ndQuanti}, we briefly recall the quantum physical formalism of the second quantization.
	A method to construct entanglement criteria is derived in Sec. \ref{subsec:FockEntanglement}.
	Finally, we apply this technique to thermal states in the system of interacting oscillators under study in Sec. \ref{subsec:FockThermal}.

\subsection{Second quantization}\label{subsec:Fock2ndQuanti}

	Let us start with a brief recapitulation of the standard technique to describe many-particle spaces mathematically.
	The Fock space $\boldsymbol{\mathcal H}$ is defined as the direct sum of all individual $N$-particle spaces,
	\begin{align}
		\boldsymbol{\mathcal H}=\bigoplus_{N=0}^\infty \mathcal H^{\otimes N}.
	\end{align}
	In general, we use the boldface notation when addressing quantities in this Fock space.
	A pure state in the Fock space, $\boldsymbol{|\psi\rangle}\in\boldsymbol{\mathcal H}$, is the direct sum of the unnormalized $N$-particle states $|\psi^{(N)}\rangle\in\mathcal H^{\otimes N}$.
	One writes this direct sum in the vector form
	\begin{align}
		\boldsymbol{|\psi\rangle}
		=\begin{pmatrix}
			|\psi^{(0)}\rangle \\ |\psi^{(1)}\rangle \\ |\psi^{(2)}\rangle \\ \vdots
		\end{pmatrix}.
	\end{align}
	Here, $|\psi^{(N)}\rangle$ represents the $N$-particle state or, equivalently, the $N$-particle wave function, $\psi^{(N)}(x_1,\ldots, x_N)$.
	It is worth mentioning that the zero-particle component is a complex number, $|\psi^{(0)}\rangle=\psi^{(0)}\in\mathbb C=\mathcal H^{\otimes0}$, and is referred to as the vacuum contribution.
	The probability $p_N$ to have $N$ particles is given by the squared norm of the $N$-particle component,
	\begin{align}\label{eq:FockParticleProb}
		p_N=\langle \psi^{(N)}|\psi^{(N)}\rangle\geq 0.
	\end{align}
	The overall normalization of the state reads
	\begin{align}\label{eq:FockNormalization}
		\boldsymbol{\langle\psi|\psi\rangle}=\sum_{N=0}^\infty\langle\psi^{(N)}|\psi^{(N)}\rangle=\sum_{N=0}^\infty p_N=1.
	\end{align}

	A linear operator $\boldsymbol{\hat L}:\boldsymbol{\mathcal H}\to\boldsymbol{\mathcal H}$ is described through its components $\hat L^{(M,N)}$, which map an $M$-particle state to an $N$-particle state,
	\begin{align}\label{eq:FockTestOp}
	\begin{aligned}
		\boldsymbol{\hat L}
		=&\begin{pmatrix}
			\hat L^{(0,0)} & \hat L^{(0,1)} & \hat L^{(0,2)} & \cdots \\
			\hat L^{(1,0)} & \hat L^{(1,1)} & \hat L^{(1,2)} & \cdots \\
			\hat L^{(2,0)} & \hat L^{(2,1)} & \hat L^{(2,2)} & \cdots \\
			\vdots & \vdots & \vdots & \ddots
		\end{pmatrix}.
	\end{aligned}
	\end{align}
	It is worth mentioning that an operator is a Hermitian one if and only if $\hat L^{(M,N)\dag}=\hat L^{(N,M)}$ for all $M,N\in\mathbb N$, and it is a block-diagonal operator if and only if $\hat L^{(M,N)}=0$ for all $M\neq N$.
	Let us consider some examples of operators which are important for our following considerations.

	The density operator of a mixed state is an operator $\boldsymbol{\hat\rho}$ in the Fock space which describes a convex combination of pure states $\boldsymbol{|\psi\rangle\langle\psi|}$, where
	$\boldsymbol{\langle\psi|}=\begin{pmatrix}
		 \langle\psi^{(0)}| & \langle \psi^{(1)}| & \cdots &
	\end{pmatrix}$.
	The particle-number operator $\boldsymbol{\hat N}$ has the following block-diagonal form
	\begin{align}
		\boldsymbol{\hat N}
		=&\begin{pmatrix}
			0 & 0 & 0 & \cdots \\
			0 & \hat 1 & 0 & \cdots \\
			0 & 0 & 2\,\hat 1^{\otimes2} & \cdots \\
			\vdots & \vdots & \vdots & \ddots
		\end{pmatrix}
		=\bigoplus_{N=0}^\infty \left(N\, \hat 1^{\otimes N}\right),
	\end{align}
	using the single-particle identity operator $\hat 1$.
	The expectation value $\overline N=\boldsymbol{\langle\hat N\rangle_{\hat \rho}}$ is the mean particle number of the state $\boldsymbol{\hat \rho}$.
	Note that the particle number should not be confused with the excitation number.
	Another block-diagonal operator is the Hamiltonian of the system under study,
	\begin{align}\label{eq:FockHamiltonian}
		\boldsymbol{\hat H}=\bigoplus_{N=0}^\infty \hat H^{(N)},
	\end{align}
	where $\hat H^{(N)}$ denotes the $N$-particle Hamiltonian in Eq. \eqref{eq:Hamiltonian} with an additional superscript ``$(N)$'' for indicating the particle number.
	Note that for $N=0$, the sums that define $\hat H^{(N)}$ are empty, which yields $\hat H^{(0)}=0$.

\subsection{Entanglement conditions}\label{subsec:FockEntanglement}

\subsubsection{Separable states in the Fock space}

	To formulate entanglement conditions, the considered notion of separability has to be defined.
	For simplicity, we restrict ourselves to the case of full separability.
	Hence, a pure separable state is the direct sum of $N$-particle product states,
	\begin{align}\label{eq:FockSeparable}
		\boldsymbol{|\psi}^\mathrm{(sep)}\boldsymbol{\rangle}
		=&\begin{pmatrix}
			\psi^{(0)} \\ |\psi_1^{(1)}\rangle \\ |\psi_1^{(2)}\rangle\otimes|\psi_2^{(2)}\rangle \\ |\psi_1^{(3)}\rangle\otimes|\psi_2^{(3)}\rangle\otimes|\psi_3^{(3)}\rangle \\ \vdots
		\end{pmatrix}
		=\bigoplus_{N=0}^\infty |\psi^{(N, \mathrm{sep})}\rangle,
	\end{align}
	where $|\psi^{(N, \mathrm{sep})}\rangle=|\psi_1^{(N)}\rangle\otimes\dots\otimes|\psi_N^{(N)}\rangle$; cf. Eq. \eqref{eq:FullSep}.
	In other words, $\boldsymbol{|\psi}^\mathrm{(sep)}\boldsymbol{\rangle}$ is factorizable for each individual particle number $N>1$.
	Analogously to the case of a fixed particle number [Eq. \eqref{eq:FullSepMixed}], a mixture of pure states yields the notion of mixed separable states in Fock spaces.

\subsubsection{Construction of entanglement criteria}

	Now, we introduce a method to construct entanglement criteria of the form \eqref{eq:EntCrit} in Fock spaces.
	This enables us to verify entanglement between particles in a system without a fixed number of particles.
	In particular, we derive the resulting separability eigenvalue equations, similar to Eq. \eqref{eq:SepEvalEqs}, for computing the desired bounds for separable states.
	Since similar derivation can be found in Refs. \cite{SV09,SV11,SV13,SSV14}, let us concisely formulate the main steps only.

	From the convexity and closure property of the set of separable states and the application of the Hahn-Banach separation theorem, it follows that a state $\boldsymbol{\hat\rho}$ is entangled if and only if there exists a Hermitian operator $\boldsymbol{\hat L}$ such that
	\begin{align}\label{eq:FockEntCrit}
		\boldsymbol{\langle \hat L\rangle_{\hat\rho}}<\lambda^\mathrm{(sep)}_{\min},
	\end{align}
	where $\lambda^\mathrm{(sep)}_{\min}$ is the minimal expectation value of $\boldsymbol{\hat L}$ for separable states.

	The latter bound is attained for pure states \eqref{eq:FockSeparable}, which are the extremal points of the set of all separable states.
	Thus, $\lambda^\mathrm{(sep)}_{\min}$ can be obtained from the minimization of
	\begin{align}\label{eq:FockOpt}
		\lambda^\mathrm{(sep)}=\boldsymbol{\langle\psi}^\mathrm{(sep)}\boldsymbol{|\hat L|}\boldsymbol{\psi}^\mathrm{(sep)}\boldsymbol{\rangle}
	\end{align}
	subjected to the constraint of normalization [Eq. \eqref{eq:FockNormalization}],
	\begin{align}\label{eq:FockNormal}
		\boldsymbol{\langle\psi}^\mathrm{(sep)}\boldsymbol{|\psi}^\mathrm{(sep)}\boldsymbol{\rangle}-1\equiv0.
	\end{align}
	In addition, we also include another restriction.

	As we mentioned initially, we do not restrict ourselves to a single particle number $N$.
	Yet, let us assume a given mean particle number $\overline N$, which allows for arbitrary fluctuations of the particle number as long as $\overline N$ remains constant.
	Hence, the second constraint is
	\begin{align}\label{eq:FockMeanN}
		\boldsymbol{\langle\psi}^\mathrm{(sep)}\boldsymbol{|\hat N|}\boldsymbol{\psi}^\mathrm{(sep)}\boldsymbol{\rangle}-\overline{N}\equiv0.
	\end{align}
	In the recent work \cite{SRLR17}, additional constraints have been also used to formulate and apply so-called ``ultrafine'' entanglement witnesses for other systems.

	In order to perform the optimization \eqref{eq:FockOpt} under the constraints \eqref{eq:FockNormal} and \eqref{eq:FockMeanN}, we can apply the method of Lagrange multipliers, which are labeled $\mu_1$ for Eq. \eqref{eq:FockNormal} and $\mu_{\overline N}$ for Eq. \eqref{eq:FockMeanN}.
	Similarly to the approach in Ref. \cite{SSV14}, this optimization over all $\langle \psi^{(N)}_j|$ directly yields our generalized separability eigenvalue equations for an operator $\boldsymbol{\hat L}$ in Eq. \eqref{eq:FockTestOp} as
	\begin{align}\nonumber
		&\sum_{M}
		\left(\bigotimes_{i<j}\langle\psi_{i}^{(N)}|\right) \left(\bigotimes_{i>j}\langle\psi_{i}^{(N)}|\right)
		\hat L^{(N,M)}
		\left(\bigotimes_{i}|\psi_{i}^{(M)}\rangle\right)
		\\=&\left(\mu_1+\mu_{\overline N} N\right)\left(\prod_{i\neq j}\langle\psi_{i}^{(N)}|\psi_{i}^{(N)}\rangle\right)|\psi_j^{(N)}\rangle,
		\label{eq:FockSepEvalEqs}
	\end{align}
	where $j,N\in\mathbb N$ and $1\leq j\leq N$ \cite{comment1}.
	Applying $\langle\psi_j^{(N)}|$ to the Fock separability eigenvalue equations \eqref{eq:FockSepEvalEqs}, summing over $N$, and using Eqs. \eqref{eq:FockOpt}, \eqref{eq:FockNormal}, and \eqref{eq:FockMeanN}, we find
	\begin{align}\label{eq:FockSepEval}
		\lambda^\mathrm{(sep)}=\mu_1+\mu_{\overline N}\overline N.
	\end{align}
	From this relation and using the minimal Fock separability eigenvalue \eqref{eq:FockSepEval}, we can compute the desired bound for the entanglement condition \eqref{eq:FockEntCrit}.

\subsubsection{Application to block-diagonal operators}

	The Fock separability eigenvalue equations \eqref{eq:FockSepEvalEqs} are obviously more complex than those in Eq. \eqref{eq:SepEvalEqs} for a fixed particle number $N$.
	For simplifying this problem, we focus on block-diagonal operators.
	In particular, the Hamiltonian in Eq. \eqref{eq:FockHamiltonian} is considered.
	In this case, $\boldsymbol{\hat L}=\boldsymbol{\hat H}$, Eq. \eqref{eq:FockSepEvalEqs} reduces to
	\begin{align}
	\begin{aligned}
		&\hat H^{(N)}_{\psi_{1}^{(N)},\ldots,\psi_{j-1}^{(N)},\psi_{j+1}^{(N)},\ldots,\psi_{N}^{(N)}}
		|\psi^{(N)}_j\rangle
		\\
		=&(\mu_1+\mu_{\overline N}N)\left(\prod_{i\neq j}\langle\psi_{i}^{(N)}|\psi_{i}^{(N)}\rangle\right)|\psi_j^{(N)}\rangle,
	\end{aligned}
	\end{align}
	for all $N$ and $j$ and using the reduced operators as defined in Eq. \eqref{eq:PartRedOp}.
	Hence, for such block-diagonal operators, the problem of solving Eq. \eqref{eq:FockSepEvalEqs} reduces to solving the separability eigenvalue equations \eqref{eq:SepEvalEqs} for all $N$ individually.

	The analytical solutions for the Hamiltonian under study are extensively discussed in the previous sections and Appendix \ref{app:FullSepEvalEqsHamiltonian}.
	Thus, we get for each $N$ the minimal separability eigenvalues
	\begin{align}\label{eq:Intersect}
		\mu_1+\mu_{\overline N}N=\frac{\hbar\Omega}{2} N\sqrt{1+(N-1)\mathcal R}=\mathcal E_{\min}^{(N,\mathrm{sep})}.
	\end{align}
	The norm of the $N$-particle component can be identified with the probability \eqref{eq:FockParticleProb} to have $N$ particles.
	Therefore, the Fock separability eigenvalue \eqref{eq:FockSepEval} for the block-diagonal operator $\boldsymbol{\hat H}$ is the convex combination
	\begin{align}\label{eq:Econvex}
		\mathcal E^{(\mathrm{sep})}=\sum_{N=0}^\infty \mathcal E_{\min}^{(N,\mathrm{sep})} p_N.
	\end{align}
	Moreover, the constraints \eqref{eq:FockNormal} and \eqref{eq:FockMeanN} rewrite as
	\begin{align}\label{eq:ConstaintedBD}
		1\equiv\sum_{N=0}^\infty p_N
		\text{ and }
		\overline N\equiv\sum_{N=0}^\infty N p_N,
	\end{align}
	respectively.

	To apply the entanglement criterion \eqref{eq:FockEntCrit} for $\boldsymbol{\hat L}=\boldsymbol{\hat H}$, the desired bound $\lambda_{\min}^\mathrm{(sep)}=\mathcal E_{\min}^\mathrm{(sep)}$ has to be obtained from the general solutions in Eq. \eqref{eq:Econvex} by minimizing over the probabilities $p_N$.
	A generalized version of this convex minimization problem is solved in Appendix \ref{app:ConvMinProbl}.
	In particular, it is shown that we can take $p_N=0$ for all $N$ except for the consecutive integers $\lfloor \overline{N}\rfloor$ and $\lfloor \overline{N}\rfloor+1$ for which holds $\lfloor \overline{N}\rfloor\leq \overline{N}<\lfloor \overline{N}\rfloor+1$.
	From the solution of the linear problem in Eq. \eqref{eq:ConstaintedBD}, we then get the probabilities $p_{\lfloor\overline N\rfloor}$ and $p_{\lfloor\overline N\rfloor+1}$, and we can finally conclude
	\begin{align}\nonumber
		\mathcal E_{\min}^\mathrm{(sep)}
		=&\frac{\hbar\Omega}{2}\Big(
			(\lfloor\overline N\rfloor+1-\overline N)\lfloor\overline N\rfloor\sqrt{1+(\lfloor\overline N\rfloor-1)\mathcal R}
		\\&+
			(\overline N-\lfloor\overline N\rfloor)(\lfloor\overline N\rfloor+1)\sqrt{1+\lfloor\overline N\rfloor\mathcal R}
		\Big).\label{eq:FockMinEnergySolution}
	\end{align}
	In summary, $\mathcal E_{\min}^\mathrm{(sep)}$ in Eq. \eqref{eq:FockMinEnergySolution} is the minimal energy of the system which can be attained for a separable state in the Fock space with a mean particle number $\overline{N}$.

\subsubsection{Example: Entanglement of the ground state}

	One can also obtain the (standard) ground state of the Hamiltonian $\boldsymbol{\hat H}$, conditioned to mean particle number $\overline N$, via the eigenvalue problem.
	Analogously to the algebra performed for separable states, this yields the minimal energy which can be attained for arbitrary states as
	\begin{align}\nonumber
		\mathcal E_{\min}
		=&\frac{\hbar\Omega}{2}\Bigg(
			(\lfloor\overline N\rfloor{+}1{-}\overline N)\left(1{+}(\lfloor\overline N\rfloor{-}1)\sqrt{1{+}\lfloor\overline N\rfloor\mathcal R}\right)
		\\&+
			(\overline N{-}\lfloor\overline N\rfloor)\left(1{+}\lfloor\overline N\rfloor\sqrt{1{+}(\lfloor\overline N\rfloor{+}1)\mathcal R}\right)
		\Bigg).
	\end{align}
	Together with Eqs. \eqref{eq:EntVis} and \eqref{eq:FockMinEnergySolution}, this allows one to compute the maximal entanglement visibility $\mathcal V_{\max}^\mathrm{(ent)}$.

\begin{figure}[t]
	\includegraphics[width=0.38\textwidth]{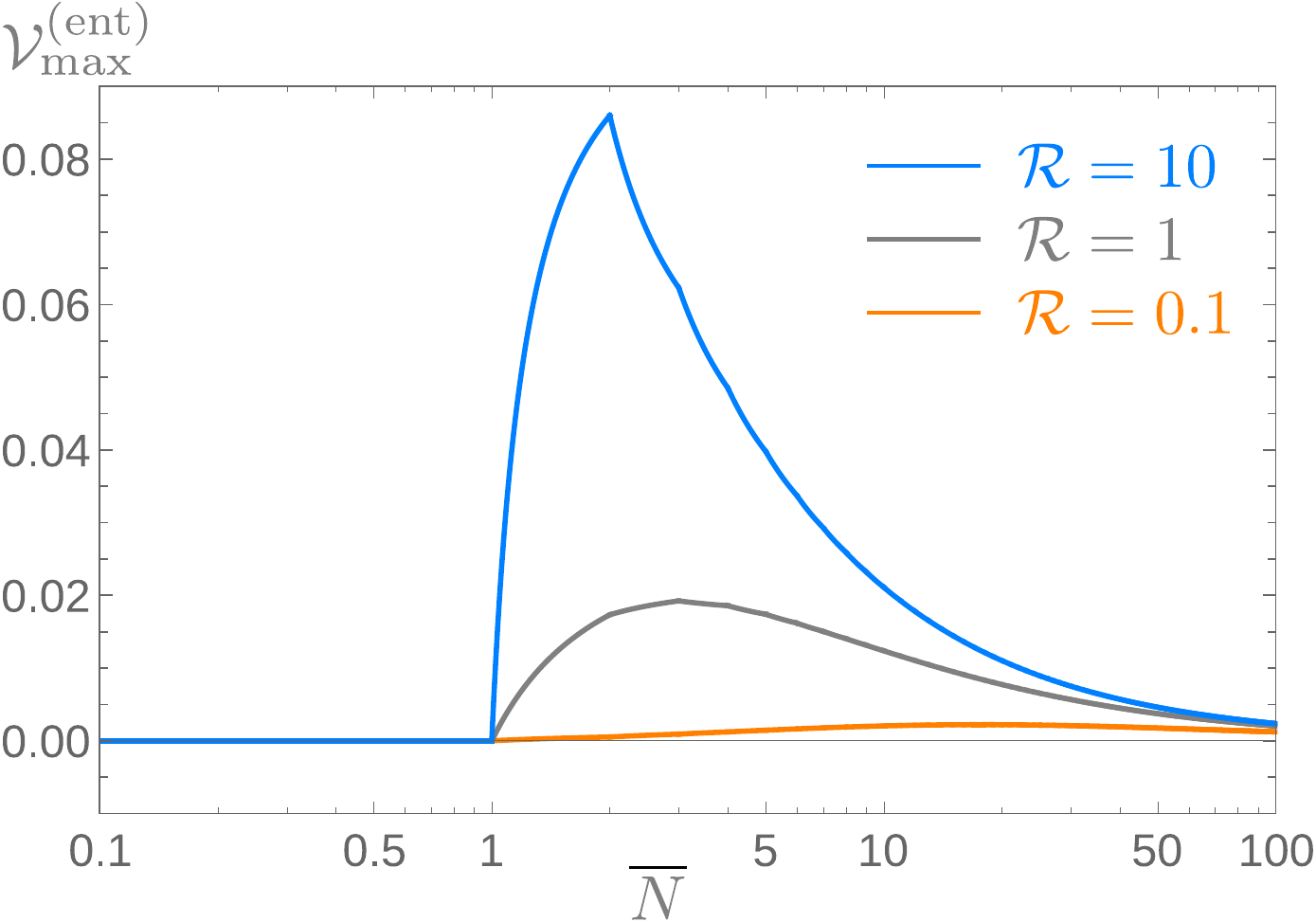}
	\caption{(Color online)
		The maximal entanglement visibility $\mathcal V^\mathrm{(ent)}_{\max}$ is depicted as a function of $\overline N$ for different $\mathcal R$ values.
		Here, the noninteger interpolation in the earlier Fig. \ref{fig:drawing2} is corrected to capture the true functional relation for mean particle numbers $\overline N$.
		For scenarios where we have a particle number less than or equal to $1$, no entanglement can be detected for any coupling for our choice of an observable, $\boldsymbol{\hat L}=\boldsymbol{\hat H}$.
	}\label{fig:MultiMeanN}
\end{figure}

	The dependency of this visibility on the mean particle number $\overline N$ is shown in Fig. \ref{fig:MultiMeanN}.
	The ground state of the considered system is entangled for any mean particle number larger than $1$.
	The nontrivial functional relation in Fig. \ref{fig:MultiMeanN} is uncovered by solving the Fock separability eigenvalue equations \eqref{eq:FockSepEval} for the Hamiltonian $\boldsymbol{\hat H}$.
	Compared to Fig. \ref{fig:drawing2}, we have the same values for the integers, i.e., $\overline N=\lfloor\overline N\rfloor$, and we do not have to make any \textit{ad hoc} interpolation for noninteger particle numbers.

\subsection{Entanglement of thermal states}\label{subsec:FockThermal}

	Let us now demonstrate how to infer entanglement of mixed states in systems with fluctuating particle numbers.
	For this reason, we consider the thermal (equilibrium) state of this interacting system.
	This thermal state is defined as \cite{K57,MS59,HWH67}
	\begin{align}\label{eq:ThermalState}
		\boldsymbol{\hat\rho}=\frac{1}{Z}
		e^{-\alpha\boldsymbol{\hat N}-\beta\boldsymbol{\hat H}},
	\end{align}
	where $\alpha=-\mu/(kT)$ (chemical potential $\mu$, temperature $T$, and Boltzmann constant $k$) and $\beta=1/(kT)$.
	Furthermore, the partition function $Z$ is given by
	\begin{align}
		Z=\boldsymbol{\mathrm{tr}}\left(e^{
			-\alpha\boldsymbol{\hat N}
			-\beta\boldsymbol{\hat H}
		}\right).
	\end{align}
	Using the previously computed energy eigenvalues of our system and $\sum_{n=0}^\infty e^{-t(n+1/2)}=(2\sinh[t/2])^{-1}$ ($t>0$), the partition function reads
	\begin{align}
		Z=&\sum_{N=0}^\infty \Gamma_N[\mathcal R],
		\text{ with}\\\nonumber
		\Gamma_N[\mathcal R]=&\frac{e^{-\alpha N}}{2\sinh[\beta u_{\mathcal E}/2]}
		\left(\frac{1}{
			2\sinh[\beta u_{\mathcal E}\sqrt{1{+}N\mathcal R}/2]
		}\right)^{N-1}.
	\end{align}
	The mean particle number $\overline N$ is given by
	\begin{align}
		\langle\boldsymbol{\hat N}\rangle_{\boldsymbol{\hat \rho}}
		=-\partial_\alpha\ln[Z]
		=\frac{1}{Z}\sum_{N=0}^\infty N\Gamma_N(\mathcal R).
	\end{align}
	We can apply $-\partial_\beta(\sinh[\beta t])^{-1}=t\coth[\beta t](\sinh[\beta t])^{-1}$, which can be used to express the mean energy of the thermal state as
	\begin{align}
		&\langle\boldsymbol{\hat H}\rangle_{\boldsymbol{\hat \rho}}
		=-\partial_\beta\ln[Z]
		\\\nonumber
		=&\frac{\hbar\Omega}{2}\Bigg(\coth\left[\frac{\hbar\Omega}{2kT}\right]
		\\\nonumber
		&+\frac{1}{Z}\sum_{N=0}^\infty (N{-}1)\sqrt{1{+}N\mathcal R}\coth\left[\frac{\hbar\Omega\sqrt{1{+}N\mathcal R}}{2k T}\right]\Gamma_N[\mathcal R]\Bigg).
	\end{align}
	Based on this mean energy and the minimal energy \eqref{eq:FockMinEnergySolution} of separable states, we can finally compute the entanglement visibility \eqref{eq:EntVis} of the thermal state for the given mean particle number $\overline N$.

\begin{figure}[t]
	\includegraphics[width=0.475\textwidth]{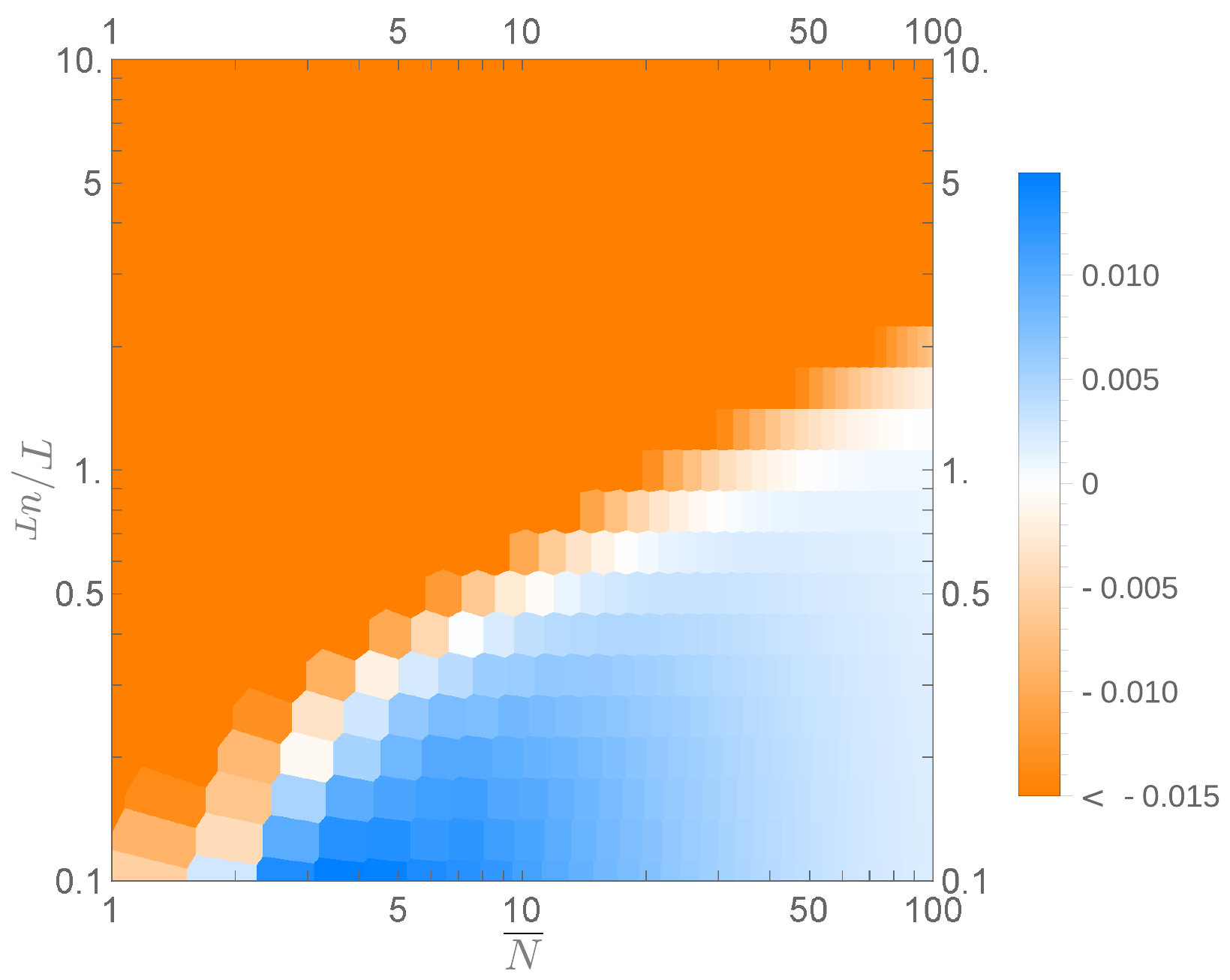}
	\caption{
		The entanglement visibility $\mathcal V^\mathrm{(ent)}$ of a thermal state as a function of $\overline N$ and $T$ for $\mathcal R=1$.
		With increasing temperature, the detection of entanglement becomes unsuccessful, $\mathcal V^\mathrm{(ent)}\leq0$.
		However, the boundary $\mathcal V^\mathrm{(ent)}=0$ increases in the depicted range with the mean number of particles.
	}\label{fig:Thermal}
\end{figure}

	The entanglement visibility of the thermal state \eqref{eq:ThermalState} is depicted in Fig. \ref{fig:Thermal}.
	The dependency of the normalized temperature $T/u_T$ [see Eq. \eqref{eq:NaturalUnits} for the definition of $u_T$] and the mean particle number $\overline N$ is shown in a range of two orders of magnitude.
	The thermal state can be entangled, which is clear when considering that a small temperature yields the (pure) entangled ground state; cf. Fig. \ref{fig:MultiMeanN}.
	In addition, with increasing temperature the entanglement visibility also decays, which also confirms the intuition that a hotter system is less entangled than a very cold one, e.g., a condensate.
	One has to keep in mind the operational meaning of the entanglement visibility $\mathcal V^\mathrm{(ent)}$ for the specific measurement of the total energy $\boldsymbol{\hat H}$, which does not exclude the possibility that entanglement persists in regions where $\mathcal V^\mathrm{(ent)}\geq0$ in Fig. \ref{fig:Thermal}.
	Applying our approach to another observable $\boldsymbol{\hat L}$ might also identify the entanglement for other regions in the $\overline N$-$T$ plane.

	Let us emphasize that entanglement of thermal states has been considered before, for example, in Refs. \cite{AW08,S14}.
	However, there the focus of attention is restricted to thermal states for a fixed number of particles---similar to our approach in Sec. \ref{sec:fixed}.
	Here, we are able to certify entanglement---already with the single observable $\boldsymbol{\hat H}$---in thermal equilibrium for temperatures $T>0$ even if the precise number of particles is undetermined.

\section{Conclusions}\label{sec:Conclusions}

	In summary, we studied entanglement properties of macroscopic systems.
	We followed two approaches.
	First, the entanglement of systems with a fixed particle number was considered.
	Second, the entanglement-detection problem was treated in systems with fluctuating particle numbers.

	We established the operational notion of entanglement visibility to quantify the detectable entanglement for a given observable.
	Moreover, we formulated a technique to construct entanglement criteria in compound systems with an unknown number of parties.
	These techniques have been applied to identify different types of multiparticle entanglement, including bipartite, full multipartite, and partial entanglement.
	In particular, we studied a system which consists of an arbitrary number of harmonic oscillators which interact with each other.

	For example, the energy spectrum restricted to separable states and the general spectrum have been compared.
	Furthermore, we determined the influence of the coupling strength on the entanglement.
	We showed for our system that the verifiable entanglement goes to zero for a macroscopic number of particles.
	This vanishing entanglement visibility has been demonstrated without introducing attenuation mechanisms or imperfections of the measurement.
	Finally, we identified entanglement of the thermal state of this correlated system.

	Let us point out that we characterized a specific physical system.
	In some second-order approximation, this system resembles the basic physics of any other ensemble of quantum particles.
	Yet, the general validity of the found dependencies of entanglement on, for example, the particle number requires additional research beyond the detailed and analytical studies presented here.
	Our derived methods can be, in principle, applied to other scenarios of interacting particles and may serve as a starting point for such investigations.

\begin{acknowledgments}
	The project leading to this application has received funding from the European Union's Horizon 2020 research and innovation programme under grant agreement No. 665148 (QCUMbER).
\end{acknowledgments}

\appendix

\section{Hermite functions}\label{app:HermiteFct}

	Let us recall some basic properties of Hermite functions.
	The Hermite functions $h^{(n)}$ are defined as
	\begin{align}
		h^{(n)}(\xi)=
		\frac{1}{\sqrt{2^nn!\sqrt\pi}}
		\left(\xi-\partial_\xi\right)^n e^{-\xi^2/2},
	\end{align}
	for $n\in\mathbb N$.
	Two examples are $h^{(0)}(\xi)=\pi^{-1/4}\exp(-\xi^2/2)$
	and $h^{(1)}(\xi)=\sqrt{2}\pi^{-1/4}\xi\exp(-\xi^2/2)$.
	Each element of this orthonormal basis solves a differential equation,
	\begin{align}\label{eq:DiffEqHermite}
		\frac{1}{2}\left(\xi^2-\partial_\xi^2\right)h^{(n)}(\xi)
		=\left(n+\frac{1}{2}\right)h^{(n)}(\xi).
	\end{align}

	In addition, they satisfy the recursion relations
	\begin{subequations}
	\begin{align}
		\xi h^{(n)}(\xi)=&\frac{\sqrt{n}h^{(n-1)}(\xi)+\sqrt{n+1}h^{(n+1)}(\xi)}{\sqrt{2}},
		\\
		\partial_\xi h^{(n)}(\xi)=&\frac{\sqrt{n}h^{(n-1)}(\xi)-\sqrt{n+1} h^{(n+1)}(\xi)}{\sqrt{2}}.
	\end{align}
	\end{subequations}
	The first- and second-order moments of $\xi$ and $\partial_\xi$ are
	$\langle\xi\rangle_{h^{(n)}}=\langle \partial_\xi\rangle_{h^{(n)}}=0$,
	$\langle\xi^2\rangle_{h^{(n)}}=-\langle\partial_\xi^2\rangle_{h^{(n)}}=n+1/2$,
	and $\langle\xi \partial_\xi\rangle_{h^{(n)}}=1/2=-\langle \partial_\xi \xi\rangle_{h^{(n)}}$.

	For our solutions in Appendix \ref{app:FullSepEvalEqsHamiltonian}, we need an operator with a displaced and rescaled potential,
	\begin{align}
		\hat L=-\frac{1}{2}\partial_\xi^2+\frac{r}{2}(\xi-\xi_0)^2+c,
	\end{align}
	with $r>0$.
	After a translation and a rescaling, we get
	\begin{align}
		\frac{\hat L-c}{\sqrt{r}}=-\frac{1}{2}\left(\partial_{\sqrt[4]{r}[\xi-\xi_0]}\right)^2+\frac{1}{2}\left(\sqrt[4]{r}[\xi-\xi_0]\right)^2,
	\end{align}
	which has the same form as the operator in Eq. \eqref{eq:DiffEqHermite}.
	Note that $\partial_xf(x-x_0)=\partial_{x-x_0}f(x-x_0)$.
	Therefore, the eigenvalue problem $\hat L\psi(\xi)=\lambda\psi(\xi)$ has the solutions
	\begin{align}
		\psi(\xi)=h^{(n)}\left(\sqrt[4]{r}[\xi-\xi_0]\right)
		\text{ and }
		\lambda=\sqrt{r}\left(n+\frac{1}{2}\right)+c,
	\end{align}
	for $n\in\mathbb N$.
	See also Ref. \cite{SCS99} for a more general treatment.

\section{Multipartite solutions}\label{app:FullSepEvalEqsHamiltonian}

	Here, we compute the exact solutions of the separability eigenvalue problem for the Hamiltonian \eqref{eq:Hamiltonian}, including the solutions for partial separability, in detail.
	The Hamiltonian in the natural units \eqref{eq:NaturalUnits} reads as
	\begin{align}\label{eq:HamiltonianRescaled}
		\hat\eta=\frac{\hat H}{u_\mathcal{E}}
		=\frac{1}{2}\sum_{1\leq i\leq N}\left(-\partial_{\xi_i}^2+\xi_i^2\right)
		+\frac{\mathcal R}{2}\sum_{1\leq i<j\leq N}\left(\xi_i-\xi_j\right)^2,
	\end{align}
	with $\xi_i=x_i/u_x$; $\mathcal R$ is defined in Eq. \eqref{eq:CouplingRatio}.
	The rescaled Hamiltonian \eqref{eq:HamiltonianRescaled} can be also written as
	\begin{align}
		\hat\eta=-\frac{1}{2}\nabla_{\vec{\xi}}^\mathrm{T}\nabla_{\vec{\xi}}+\frac{1+N\mathcal R}{2}\vec{\xi}^\mathrm{T}\vec{\xi}-\frac{\mathcal R}{2}\vec{\xi}^\mathrm{T}\vec{n}\vec{n}^\mathrm{T}\vec{\xi},
	\end{align}
	with $\vec{\xi}=(\xi_i)_{i=1}^N$ and a constant vector $\vec n=(1)_{i=1}^N$.

	Now, we consider a $K$-partition $\mathcal I_1{:}\cdots{:}\mathcal I_K$ of the set $\{1,\dots,N\}$, where each subset $\mathcal I_j$ consists of $N_j$ elements.
	We define $\vec{\xi}_j=(\xi_i)_{i\in\mathcal I_j}$ and $\vec{n}_j=(1)_{i\in\mathcal I_j}$, which yields
	\begin{align}
		\hat\eta=&
		\frac{1}{2}\sum_{i=1}^K\left(-\nabla_{\vec{\xi}_i}^\mathrm{T}\nabla_{\vec{\xi}_i}+(1+N\mathcal R)\vec{\xi}_i^\mathrm{T}\vec{\xi}_i\right)
		\\&\nonumber
		-\frac{\mathcal R}{2}\sum_{1\leq i\leq K}\vec{\xi}_i^\mathrm{T}\vec{n}_i\vec{n}_i^\mathrm{T}\vec{\xi}_i
		-\mathcal R\sum_{1\leq i<j\leq K}\vec{\xi}_i^\mathrm{T}\vec{n}_i\vec{n}_j^\mathrm{T}\vec{\xi}_j.
	\end{align}
	We may separate parts that are parallel to $\vec{n}_j$ from those that are perpendicular ($\vec{n}_j^\mathrm{T}\vec{n}_j=N_j$),
	\begin{align}
		\xi_j^{(\parallel)}=\frac{\vec{n}_j^\mathrm{T}\vec{\xi}_j}{\sqrt{N_j}}
		\text{ and }
		\vec{\xi}_j^{\,(\perp)}=\left(\mathrm{Id}_j-\frac{\vec{n}_j\vec{n}_j^\mathrm{T}}{N_j}\right)\vec{\xi}_j,
	\end{align}
	with the $N_j\times N_j$ identity matrix $\mathrm{Id}_j$.
	This allows one to bring the Hamiltonian in the form
	\begin{align}
	\begin{aligned}
		\hat\eta=&
		\frac{1}{2}\sum_{i=1}^K\left(
 		-\nabla_{ \vec{\xi}_i^{(\perp)} }^\mathrm{T}\nabla_{ \vec{\xi}_i^{(\perp)} }
 		+(1+N\mathcal R) \vec{\xi}_i^{(\perp)} {}^{\mathrm{T}}
 		\vec{\xi}_i^{(\perp)}
		\right)
		\\&+
		\frac{1}{2}\sum_{j=1}^K\left(-\partial^2_{\xi_j^{(\parallel)}}+\big(1+(N-N_j)\mathcal R\big){\xi_j^{(\parallel)}}^2\right)
		\\&-
		\mathcal R\sum_{1\leq i<j\leq K}\sqrt{N_iN_j}\xi_i^{(\parallel)}\xi_j^{(\parallel)}.
	\end{aligned}
	\end{align}

	In the following step, we analyze the reduced operator $\hat\eta_{\psi_{1},\ldots,\psi_{j-1},\psi_{j+1},\ldots,\psi_{K}}$ [Eq. \eqref{eq:PartRedOp}].
	For a clearer overview, all terms that do not depend of the remaining degree of freedom $j$ are denoted as ``const$_j$''.
	We get
	\begin{align}\label{eq:EtaOpSomeManipulations}
	\begin{aligned}
		&\hat\eta_{\psi_{1},\ldots,\psi_{j-1},\psi_{j+1},\ldots,\psi_{K}}
		\\=&
		\frac{1}{2}\left(-\nabla_{\vec{\xi}_j^{\,(\perp)}}^\mathrm{T}\nabla_{\vec{\xi}_j^{\,(\perp)}}+(1+N\mathcal R)\vec{\xi}_j^{\,(\perp)}{}^\mathrm{T}\vec{\xi}_j^{\,(\perp)}\right)
		\\&+
		\frac{1}{2}\left(-\partial^2_{\xi_j^{(\parallel)}}+\big(1+(N-N_j)\mathcal R\big){\xi_j^{(\parallel)}}^2\right)
		\\&-
		\mathcal R\sqrt{N_j}\left(\sum_{i\neq j}\sqrt{N_i} \langle \xi_i^{(\parallel)}\rangle_{\psi_i}\right)\xi_j^{(\parallel)}
		+\mathrm{const}_j.
	\end{aligned}
	\end{align}
	The $N_j-1$ degrees of freedom $\vec{\xi}_j^{\,(\perp)}$ are not influenced by the other subsystems.
	The parallel component $\xi_i^{(\parallel)}$, however, is displaced, which can be seen in the form
	\begin{align}
		\nonumber&\hat\eta_{\psi_{1},\ldots,\psi_{j-1},\psi_{j+1},\ldots,\psi_{K}}
		\\=&\nonumber
		\frac{1}{2}\left(-\nabla_{\vec{\xi}_j^{\,(\perp)}}^\mathrm{T}\nabla_{\vec{\xi}_j^{\,(\perp)}}+(1+N\mathcal R)\vec{\xi}_j^{\,(\perp)}{}^\mathrm{T}\vec{\xi}_j^{\,(\perp)}\right)
		-\frac{1}{2}\partial^2_{\xi_j^{(\parallel)}}
		\\&+\nonumber
		\frac{1{+}(N{-}N_j)\mathcal R}{2}\left(
			\xi_j^{(\parallel)}
			{-}\frac{
				\mathcal R
				\sum_{i\neq j}\sqrt{N_iN_j}\langle \xi_i^{(\parallel)}\rangle_{\psi_i^{(\parallel)}}
			}{1+(N-N_j)\mathcal R}
		\right)^2
		\\&+\mathrm{const}_j.
	\end{align}
	Hence, the solutions in terms of displaced Hermite functions (Appendix \ref{app:HermiteFct}) have the mean position
	\begin{align}
	\begin{aligned}
		\langle\xi_j^{(\parallel)} \rangle_{\psi_j^{(\parallel)}}=
		\frac{
			\mathcal R\sqrt{N_j}
			\sum_{i\neq j}\sqrt{N_i}\langle \xi_i^{(\parallel)}\rangle_{\psi_i^{(\parallel)}}
		}{1+(N-N_j)\mathcal R}
		\quad\Leftrightarrow
		\\
		\frac{1+N\mathcal R}{\sqrt{N_j}}
		\langle\xi_j^{(\parallel)} \rangle_{\psi_j^{(\parallel)}}
		=
		\mathcal R
		\sum_{i=1}^K\sqrt{N_i}\langle \xi_i^{(\parallel)}\rangle_{\psi_i^{(\parallel)}}=c,
	\end{aligned}
	\end{align}
	where the center part is independent of $j$ and defines the constant $c$.
	Thus, we have $\langle\xi_j^{(\parallel)} \rangle_{\psi_j^{(\parallel)}}=\sqrt{N_j}c/(1+N\mathcal R)$, which can be inserted into the above definition of $c$,
	\begin{align}\label{eq:cDef}
		c=\frac{\mathcal R}{1+N\mathcal R}\sum_{i=1}^K N_ic
		=\frac{N\mathcal R}{1+N\mathcal R}c,
	\end{align}
	where we used $\sum_{i=1}^K N_i=N$.
	Equation \eqref{eq:cDef} is only fulfilled if $c=0$.
	This results in $\langle\xi_j^{(\parallel)} \rangle_{\psi_j^{(\parallel)}}=0$ for all $j$.

	Taking this information and Appendix \ref{app:HermiteFct} into account, we can directly solve the eigenvalue equations \eqref{eq:SepEvalEqs} of the operator \eqref{eq:EtaOpSomeManipulations}.
	This yields the wave functions of the separability eigenvectors,
	\begin{align}\label{eq:AllSEvec}
	\begin{aligned}
		\psi^{(\mathcal I_1{:}\cdots{:}\mathcal I_K)}(\vec{\xi})=&\psi_1(\vec{\xi}_1)\cdots\psi_K(\vec{\xi}_K),\text{ with}
		\\
		\psi_j(\vec{\xi}_j)=&
		\psi_j^{(\perp)}(\vec{\xi}_j^{\,(\perp)})
		\psi_j^{(\parallel)}(\xi_j^{(\parallel)}),
		\\
		\psi_j^{(\parallel)}(\xi_j^{(\parallel)})
		=&h^{(n_j^{(\parallel)})}\left(
			\sqrt[4]{1+(N-N_j)\mathcal R}\xi_j^{(\parallel)}
		\right),
		\\
		\psi_j^{(\perp)}(\vec{\xi}_j^{\,(\perp)})
		=&h^{(\vec{n}_j^{(\perp)})}\left(
			\sqrt[4]{1+N\mathcal R}\vec{\xi}_j^{\,(\perp)}
		\right),
	\end{aligned}
	\end{align}
	where $n_j^{(\perp)}\in\mathbb N$ and $\vec{n}_j^{(\perp)}\in\mathbb N^{N_j-1}$.
	Here, $h^{(\vec{n}_j^{(\perp)})}$ is a product of $N_j-1$ (for each degree of freedom of $\vec{\xi}_j^{\,(\perp)}$) Hermite functions of the orders as defined by $\vec{n}_j^{(\perp)}$.
	Also, we get the separability eigenvalues
	\begin{align}\label{eq:AllSEval}
	\begin{aligned}
		\lambda^{\mathcal I_1{:}\cdots{:}\mathcal I_K}
		=&
		\langle\hat\eta\rangle_{\psi^{(\mathcal I_1{:}\cdots{:}\mathcal I_K)}}
		\\=&\sum_{j=1}^K\sqrt{1+N\mathcal R}\left(|\vec{n}_j^{(\perp)}|_1+\frac{N_j-1}{2}\right)
		\\&+\sum_{j=1}^K\sqrt{1+(N-N_j)\mathcal R}\left( n_j^{(\parallel)}+\frac{1}{2}\right),
	\end{aligned}
	\end{align}
	where $|\vec{n}_j^{(\perp)}|_1$ denotes the $1$-norm of $\vec{n}_j^{(\perp)}$.

	For a trivial partition, $K=1$ or $\mathcal I_1=\{1,\dots, N\}$, the separability eigenvalue equations coincide with the standard eigenvalue equations of the rescaled Hamiltonian \eqref{eq:HamiltonianRescaled}.
	Thus, the eigenfunctions and eigenvalues can be obtained as special cases of Eqs. \eqref{eq:AllSEvec} and \eqref{eq:AllSEval},
	\begin{align}\label{eq:AllEvec}
		\psi(\vec{\xi})
		=&h^{(n^{(\parallel)})}\left(
			\xi_j^{(\parallel)}
		\right)
		h^{(\vec{n}^{(\perp)})}\left(
			\sqrt[4]{1+N\mathcal R}\vec{\xi}^{(\perp)}
		\right)
	\end{align}
	and
	\begin{align}\label{eq:AllEval}
		\lambda
		=&\sqrt{1+N\mathcal R}\left(|\vec{n}^{(\perp)}|_1+\frac{N{-}1}{2}\right)
		+\left( n^{(\parallel)}+\frac{1}{2}\right).
	\end{align}
	Full separability, $K=N$ or $\mathcal I_j=\{j\}$, can be also directly concluded.
	In this case, one should point out that $N_j=1$ and thus $\xi^{(\parallel)}_j=\xi_j$ and $\vec{\xi}^{(\perp)}_j$ is a zero-dimensional vector (i.e., a vanishing contribution).
	We obtain from Eqs. \eqref{eq:AllSEvec} and \eqref{eq:AllSEval} the following:
	\begin{align}\label{eq:AllfullSEvec}
	\begin{aligned}
		\psi^\mathrm{(sep)}(\vec{\xi})=&\prod_{j=1}^N h^{(n_j)}\left(
			\sqrt[4]{1+(N-1)\mathcal R}\xi_j
		\right),
	\end{aligned}
	\end{align}
	and
	\begin{align}\label{eq:AllfullSEval}
	\begin{aligned}
		\lambda^\mathrm{(sep)}
		=\sum_{j=1}^N\sqrt{1+(N-1)\mathcal R}\left( n_j+\frac{1}{2}\right),
	\end{aligned}
	\end{align}
	where we skipped the superscript ``$(\parallel)$''.

\section{Minimization and convex functions}\label{app:ConvMinProbl}

	Let us show that the separability eigenvalue \eqref{eq:Econvex} constrained to a mean particle number $\overline N\notin\mathbb N$ is attained for a mixture of $N\in\{\lfloor \overline N\rfloor,\lceil \overline N\rceil\}$, using the floor function ($\lfloor x\rfloor=\max\{n\in\mathbb N: n\leq x\}$) and the ceiling function ($\lceil x\rceil=\min\{n\in\mathbb N: n\geq x\}$).
	For $\overline N\in\mathbb N$, we have $\overline N=N$.
	We prove a more general statement.

	Suppose the following:
	$f$ is a convex function, $f[px+(1-p)]\leq pf[x]+(1-p)f[y]$ for all $0\leq p\leq 1$;
	$f$ is not an affine function, $f[x]\neq t_1x +t_0$ for all $t_1$ and $t_0$;
	and $a\leq a'\leq x\leq b'\leq b$.
	We can always write
	\begin{align}\label{eq:ConvexBounds}
		a'=q a+(1-q) b
		\text{ and }
		b'=r a+(1-r) b,
	\end{align}
	with $0\leq q\leq r\leq 1$ to ensure $0<b'-a'=(q-r)(b-a)$.

	It is obvious that $f$ and any affine function can only have up to two points in common (intersection with a secant).
	Let $g$ and $g'$ be two affine functions for which holds $f[a]=g[a]$ and $f[a']=g'[a']$ and analogously for $b$ and $b'$.
	Thus, we can write
	\begin{align}
	\begin{aligned}
		g[x]&=f[a]\frac{b-x}{b-a}+f[b]\frac{x-a}{b-a},
		\\
		g'[x]&=f[a']\frac{b'-x}{b'-a'}+f[b']\frac{x-a'}{b'-a'}.
	\end{aligned}
	\end{align}
	Inserting the convex decomposition of $a'$ and $b'$ in Eq. \eqref{eq:ConvexBounds} and using the convexity of $f[x]$ itself (at the points $x=a'$ and $x=b'$), one directly finds that
	\begin{align}\label{eq:LinInterpolBounds}
		g'[x]\leq g[x],
	\end{align}
	where the decomposition $x=px+(1-p)x$ is helpful.

	Inequality \eqref{eq:LinInterpolBounds} states that the minimal energy (convex function $f$) is attained for the two integer values (intersection points $a'$ and $b'$) that are closest to $\overline N$ (argument $x$).
	In detail, more narrow bounds $a'$ and $b'$ result in the interpolation of $f[x]$ with a better and smaller value $g[x]$ compared to $g'[x]$ for any $a\leq a'$ and $b\geq b'$.
	Also note that any convex combination with more than two elements yields a set $\mathrm{conv}\{(x,f[x]):x\in\mathbb N \}$ which is bounded by secants.



\begin{thebibliography}{99}
\bibitem{comment2}
		For example, one partially reduced operator of $\hat L=\sum_s \hat A_{s}\otimes\hat B_{s}$ in a bipartite case $\mathcal H_A\otimes\mathcal H_B$ takes the form $\hat L_a=\sum_s\langle a|\hat A_{s}|a\rangle \hat B_{s}$.
		It depends on $|a\rangle\in\mathcal H_A$ and maps a vector $|b\rangle$ onto $\hat L_a|b\rangle\in\mathcal H_B$.
		See also Refs. \cite{SV09,SV13}.
\bibitem{comment3}
		We have $\kappa_{\min}^\mathrm{(sep)}=\mu+\nu\lambda_{\min}^\mathrm{(sep)}$ for an $N$-particle operator $\hat K=\mu\hat 1^{\otimes N}+\nu\hat L$ ($\mu,\nu\in\mathbb R$ and $\nu>0$); cf. Ref. \cite{SV13}.
		This holds similarly for standard eigenvalues, $\kappa_{\min}=\mu+\nu\lambda_{\min}$, and expectation values, $\langle\hat K\rangle_{\hat\rho}=\mu+\nu\langle\hat L\rangle_{\hat\rho}$.
\bibitem{comment1}
		For an operator $\hat X:\mathcal H_D\to\mathcal H_A\otimes\mathcal H_B\otimes\mathcal H_C$, the notation $\langle \alpha|\langle\gamma|\hat X|\delta\rangle$ defines a linear map $\mathcal H_B^\ast\to\mathbb C$ for the given vectors $|\alpha\rangle\in\mathcal H_A$, $|\gamma\rangle\in\mathcal H_C$, and $|\delta\rangle\in\mathcal H_D$.
		It maps $\langle\beta|\in\mathcal H_B^\ast$ onto the complex number $(\langle\alpha|\otimes\langle\beta|\otimes\langle\gamma|) \hat L|\delta\rangle$.
\end{thebibliography}
\end{document}